\documentclass[a4paper,11pt]{article}
\pdfoutput=1
\usepackage{jheppub} 
\usepackage[bottom]{footmisc}
\usepackage{amssymb}
\usepackage{amsmath}
\usepackage{amsthm}
\usepackage[usenames,dvipsnames]{xcolor}
\usepackage{epsfig}
\usepackage{dcolumn}
\usepackage{tikz}
\usetikzlibrary{shapes.geometric, arrows}
\usepackage{upgreek}
\usepackage{setspace}
\usepackage{subfig}
\usepackage{enumitem}
\usepackage{array,multirow,bigdelim,arydshln}
\usepackage{appendix}
\usepackage{xparse}
\usepackage{stmaryrd}
\usepackage[T1]{fontenc} 
\usepackage{mathtools}
\usepackage{physics} 
\usepackage{adjustbox}
\usepackage{multirow}
\usepackage{graphicx} 
\usepackage{float} 
\graphicspath{{./images/}}
\usepackage[nottoc]{tocbibind}
\usepackage{hyperref}
\usepackage[utf8]{inputenc}
\usepackage{CJK}
\hypersetup{
	colorlinks,
	urlcolor=Maroon,
	linkcolor=Maroon,
	citecolor=Maroon
	}

\usepackage{latexsym}
\usepackage{tikz}

\theoremstyle{plain}

\theoremstyle{definition}

\def\bea#1\eea{\begin{eqnarray}#1\end{eqnarray}}
\def\be#1\ee{\begin{equation}#1\end{equation}}
\def\ba#1\ea{\begin{align}#1\end{align}}

\usepackage{amsmath}
\usepackage{multicol}
\usepackage{bbm}
\usepackage{amsthm}
\usepackage{mathrsfs}
\usepackage{upgreek}
\usepackage{amssymb}
\usepackage{bm}
\usepackage{setspace}
\usepackage{array,multirow,arydshln}
\usepackage{bigdelim}
\usepackage{scalerel}
\usepackage{diagbox}

\usepackage{tabularx}
\usepackage{tikz}

\usetikzlibrary{shapes.geometric,arrows,arrows.meta,decorations.pathmorphing,decorations.markings,patterns}

\def\<{\langle}
\def\>{\rangle}

\usepackage[percent]{overpic}
\usepackage{multirow} 
\usepackage{slashed}

\title{Circles and Triangles, the NLSM and Tr($\Phi^3$)}

\author[a]{Nima Arkani-Hamed,}
\author[b]{Carolina Figueiredo}

\affiliation[a]{School of Natural Sciences, Institute for Advanced Study, Princeton, NJ, 08540, USA}
\affiliation[b]{Jadwin Hall, Princeton University, Princeton, NJ, 08540, USA}

\emailAdd{arkani@ias.edu}
\emailAdd{cfigueiredo@princeton.edu}

\abstract{A surprising connection has recently been made between the amplitudes for Tr($\Phi^3$) theory and the non-linear sigma model (NLSM). A simple shift of kinematic variables naturally suggested by the associahedron/stringy representation of Tr$(\Phi^3$) theory yields pion amplitudes at all loops. In this note we provide an elementary motivation and proof for this link going in the opposite direction, starting from the non-linear sigma model and discovering its formulation as a sum over triangulations of surfaces with simple numerator factors. This uses an ancient connection between ``circles'' and ``triangles'',  interpreting the equation $y = \sqrt{1 - x^2}$ both as parametrizing points on a circle as well as generating the number of triangulations of polygons. A further simplification of the numerator factors exposes them as arising from the kinematically shifted Tr($\Phi^3$) theory, and gives rise to novel tropical representations of NLSM amplitudes. The connection to  Tr$(\Phi^3)$ theory defines a natural notion of ``surface-soft limit'' intrinsic to curves on surfaces. Remarkably, with this definition, the soft limit of pion amplitudes vanishes directly at the level of the integrand, via obvious pairwise cancellations. We also give simple, explicit expressions for the multi-soft factors for tree and loop-level integrands in the limit as any number of pions are taken ``surface-soft''.}

\begin{document}

\begin{CJK*}{UTF8}{}
\CJKfamily{gbsn}
\maketitle
\end{CJK*}
\addtocontents{toc}{\protect\setcounter{tocdepth}{2}}

\numberwithin{equation}{section}

		\tikzset{
		particles/.style={dashed, postaction={decorate},
			decoration={markings,mark=at position .5 with {\arrow[scale=1.5]{>}}
		}}
	}
	\tikzset{
		particle/.style={draw=black, postaction={decorate},
			decoration={markings,mark=at position .5 with {\arrow[scale=1.1]{>}}
		}}
	}
	\def  \layersep {.6cm}

\section{Introduction}
\label{sec:Intro}
Scattering amplitudes for pions are described by the non-linear sigma model, and have been intensively studied since the 1960's. The past decade has seen a renaissance in these investigations from the on-shell perspective, giving rise to several new ways of computing NLSM amplitudes ranging from from recursion relations~\cite{RecFromSoftTheorems,RecursionJara,OnShellRecRel}, to bootstrap methods~\cite{AllEFTsCliffJara,DoubleCopyAndSoft,EFTsSoftLimits,ScalarBCJBoostrap,softBootstrap,Uniqueness2,UniquenessNLSM,UVconisderations} and other new representations~\cite{CHY,FlavorKinRepresentation}. 

Recently a surprising connection has been made between scattering amplitudes for Tr$(\Phi^3)$ theory and the non-linear sigma model (NLSM), via a simple shift of kinematic data~\cite{zeros,deltaNLSM}. Defining the planar variables for a given color-ordering $X_{i,j} = (p_i + p_{i+1} + \cdots p_{j-1})^2$, we begin with Tr$(\Phi^3)$ amplitudes at tree-level and shift 
\begin{equation}
X_{e,e} \to X_{e,e} + \delta, \quad X_{o,o} \to X_{o,o} - \delta,
\end{equation}
where $e,o$ are even and odd indices respectively. Then the leading low-energy amplitudes for $X_{i,j} \ll \delta$ are those of the NLSM, with the full amplitude providing an unusual UV completion of the NLSM. A simple generalization averaging over even/odd assignments extends to all loop orders~\cite{deltaNLSM}. 

This connection is both motivated and made obvious starting from the formulation of Tr($\Phi^3$) amplitudes in terms of associahedra and their stringy avatars~\cite{ABHY,GiulioClusters,StringForms,StringyCanonicalForms}. But if we instead begin with the NLSM, what would make us suspect any link of this sort? Why should the amplitudes for a Lagrangian with infinitely many even-point interactions have anything in common with those coming from a simple sum over cubic diagrams? We will begin this note by giving a  nice answer to this question.  

The NLSM Lagrangian is
\begin{equation}
    {\cal L} = \frac{1}{2} {\rm Tr} \left(\partial_\mu U^\dagger \partial^\mu U \right) , \, \, \, U^\dagger U = 1,
\end{equation}
where we work in units such that the pion decay constant $f_\pi = 1$. There are several ways of parametrizing the solution of the $U^\dagger U = 1$ constraints, a common one is the  ``minimal'' or ``square-root'' parametrization  
\begin{equation}
U = \Phi + i \sqrt{1 - \Phi^2},
\end{equation}
which is the same as the parametrization of points on the unit circle in the complex plane $z = x + i \sqrt{1 - x^2}$. 

Now, there is a famous connection between ``circles'' and `` triangulations'': the same square-root generates the Catalan numbers $C_n$ that count the number of triangulations of an $(n+2)$-gon. We have that $C_0 = 1$ by convention, $C_1=1$, $C_2=2$ for the two ways of triangulating a square, $C_3 = 5$ for the five ways of triangulating a pentagon, $C_4 = 14$ and so on. The definition of $C_n$ as counting triangulations makes it obvious that it satisfies the quadratic recursion $C_{n+1} = \sum_k C_k C_{n-k}$. Defining the generating function 
\begin{equation}
F(t) = \sum_n C_n t^n = 1 + t + 2t^2 + 5 t^3 + 14 t^4 + \cdots 
\end{equation}
this recursion implies that $F(t) = t F(t)^2 + 1$, and solving this quadratic equation gives 
\begin{equation}
F(t) = \frac{(1 - \sqrt{1 - 4 t})}{2 t}.
\end{equation}

We can make this ``circle/triangulations'' relation completelty explicit by writing
\begin{equation}
\sqrt{1 - x^2} = 1 - \frac{x^2}{2} \left[1 + \left( \frac{x}{2} \right) + 2 \left( \frac{x}{2} \right)^2 + 5 \left( \frac{x}{2} \right)^3 + + 14 \left(\frac{x}{2} \right)^4 + \cdots \right]. 
\end{equation}

This connection between ``circles'' and ``triangulations'' is ubiquitous in mathematics and physics, for instance, is the same square-root that relates the Wigner semi-circle law for the eigenvalue distribution of Gaussian random matrices $\Phi$ to the diagrammatic computation of $\langle \Phi^m \rangle$ where the Catalan numbers occur in counting nesting bracketings, as in Wigners' original computation~\cite{WignerSemiCircle}. 

As we will see, this link allows us to give a representation of the $2n$-particle NLSM amplitudes as a sum over triangulations of a $2n$-gon,  with simple numerator factors associated with chords contained in ``even-angulations'' that underlie full triangulations. We will then see that a change to an even simpler rule but with ``even-even, odd-odd'' asymmetry, reproduces the shifted Tr$(\Phi^3)$ theory. The two rules are equivalent due to certain transparent cancellations that take place due to the interesting relative minus signs inherited from $X_{e,e} \to X_{e,e} + \delta, X_{o,o} \to X_{o,o} - \delta$ kinematic shifts. This final formulation of the NLSM as a sum over triangulations also gives us an extremely simple and efficient tropical representation of NLSM amplitudes. 

In addition to its striking simplicity, this formulation of the NLSM gives a very direct understanding of the Adler zero~\cite{AdlerZero} for pion amplitudes at tree-level, arising from combinatorially obvious pairings of triangulations that cancel in the soft limit. At loop-level, it has recently been noted \cite{JaraSoftTheorem} that the Adler zero is not present at the level of the standard loop integrand, even taking external bubble legs off-shell in order to avoid the usual $1/0$ ambiguities associated with bubbles. Instead in the soft-limit, the integrand derived from the shifted Tr$(\Phi^3)$ theory reduces to nice objects multiplying scale-less integrands that integrate to zero.  But surprisingly, we will show that the connection to surfaces defines a slightly different and more natural notion of soft limit, generalizing subtly away from the obvious notion in momentum space. With this new definition of ``surface-soft'' limit, we find the Adler zero directly at the level of the integrand! This holds not only for the NLSM but for the shifted Tr$(\Phi^3)$ amplitudes at any finite value of the deformation $\delta$. There is a deep and simple conceptual origin for these statements, naturally formulated in the world of surfaces and the binary $u$-variables, that also makes the extension to multi-soft limits obvious \cite{upcoming}. Nonetheless, we will describe a direct combinatorial understanding of the Adler zeros for the NLSM loop integrand in the ``surface-soft'' limit, arising from essentially the same pairwise cancellations seen at tree-level. 

\section{The Catalan representation: NLSM from Triangulations}
\begin{figure}[t]
    \centering
\includegraphics[width=0.4\textwidth]{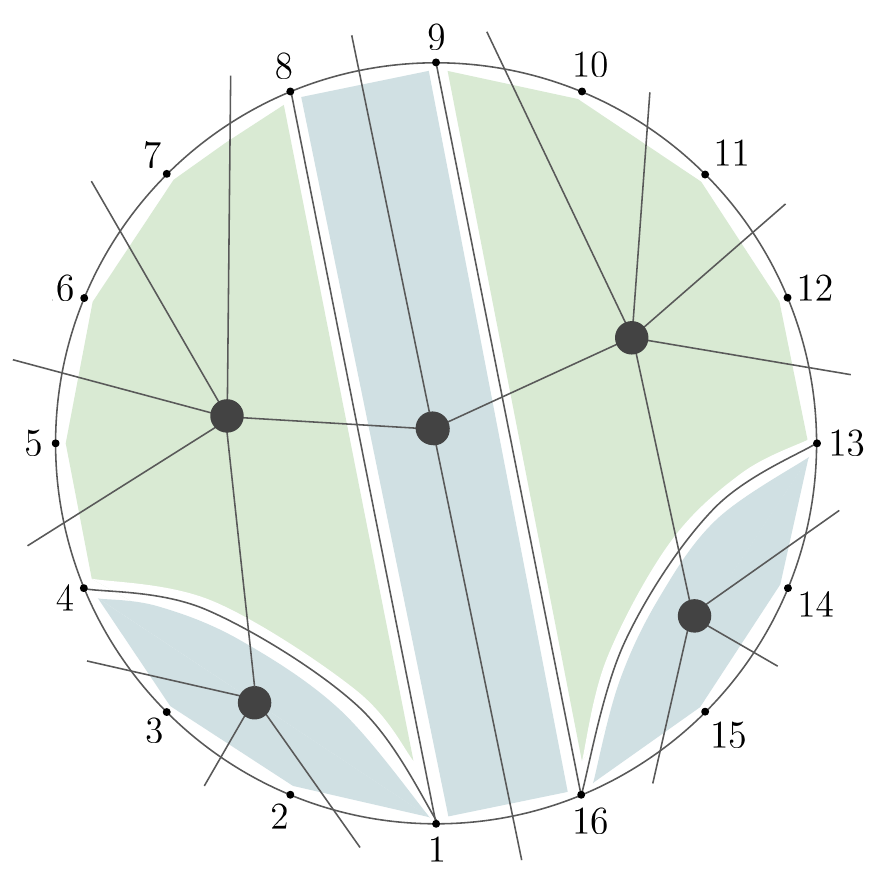}
    \caption{An even-angulation of the $2n=16$-gon. The shaded blue regions are $4$-gons and the shaded green are $6$-gons. The corresponding dual Feynman graph is shown, gluing together $4$- and $6$-valent vertices.}
    \label{fig:dualgraph}
\end{figure}

Using the minimal parametrization $U=\Phi + i \sqrt{1 - \Phi^2}$ in the NLSM Lagrangian, the even $(2m + 2)$-point interaction vertices ${\cal L}^{(2m+2)}$ can nicely be written as \cite{JaraNLSM}:
\begin{equation}
{\cal L}^{(2m + 2)}\left(X_{i, j}\right)= \frac{1}{2} \sum_{k=0}^{m-1} C_{k-1} C_{m-k} \sum_{i=1}^{2 m+2} X_{i, i+2 k},
\label{eq:CatalanInt}
\end{equation}
where we see the appearance of the Catalan numbers, coming from the expansion of the square-root as alluded to above. Note that in this representation the (off-shell) interactions only depend on the internal $X_{i,j}$ variables $X_{i,j} = (k_i + k_{i+1} + \cdots k_{j-1})^2$, where $k_i$ are the off-shell momenta, but not on $X_{i,i+1} = k_i^2$, which are non-zero off-shell. This means that the contact term of the $(2m+2)$-point amplitude is exactly given by ${\cal L}^{(2m + 2)}$. To get the $2n$ tree amplitude, we simply sum over all ``even-angulations'' ${\cal E}$ of the $2n$-gon--subdividing the $2n$-gon into non-overlapping $2m$-gons, dual to Feynman diagrams made from even-valence interactions as in fig. \ref{fig:dualgraph}, weight each $2m + 2$-gon with the factor ${\cal L}^{(2m+2)}$ and multiply by all the poles $1/X$ for the chords in ${\cal E}$,  

We will now give a simple combinatorial interpretation of this contact term that will also lead to a simple and natural expression for the full $2n$-particle NLSM amplitude given as a sum over all triangulations of the $2n$-gon. Other interesting representations of the NLSM as a sum over cubic graphs have been given in \cite{cubicnum1,cubicnum2,cubicnum3}.

The first feature to highlight is that the variables appearing in ${\cal L}^{(2m + 2)}$ are the $X_{i,j}$'s with $|i-j|$ even -- these are precisely associate with propagators that isolate odd-point interactions. Given that odd-point pion amplitudes vanish, these variables never occur as poles: all the poles correspond to $X_{e,o}$ variables. 

Let's now interpret the Catalan factors multiplying a particular $X_{i,i+2k}$. For this it is useful to represent the momenta entering the $(2m+2)$-point interaction using a momentum $(2m+2)$-gon, in which case ${\cal L}^{(2m+2)}$ is a function of the chords, $X_{i,j}$, of the $(2m+2)$-gon, with $(i,j)$ both even/odd. Therefore chord $X_{i,i+2k}$ divides the $(2m+2)$-gon into a $(2k+1)$-gon and a $(2m-2k+3)$-gon (both odd-gons as expected). Now if we ask how many ways there are of triangulating the $(2k+1)$-gon with \textbf{only} $X_{e,e}/X_{o,o}$ chords, we obtain precisely $C_{k-1}$. We can prove this easily by noting that any such triangulation of the $(2k+1)$-gon will have to contain the chords $X_{i,i+2},X_{i+2,i+4},\cdots,X_{i+2k-2,i+2k}$, which bounds a $(k+1)$-gon where all the vertices are even/odd, depending on the parity of $i$. Therefore the number of $(e,e)/(o,o)$ triangulations $(2k+1)$-gon is simply the number of triangulations of this inner $(k+1)$-gon that is precisely $C_{k-1}$. Following exactly the same argument we obtain $C_{m-k}$ for the $(2m-2k+3)$-gon. So the factor multiplying a given $X_{i,i+2k}$ corresponds to the product of the number of ways of triangulating the two lower-point problems, using \textbf{exclusively} $X_{e,e}/X_{o,o}$ chords. 

Using this interpretation, we can construct the Feynman rules that give the full $2n$-point pion tree amplitudes. We consider all ``even-angulations'', ${\cal E}$, of the $2n$-gon, $i.e.$ tilings that divide it into a collection of even-point polygons. The chords entering the even-angulation are all $(e,o)$, giving us the propagators $\prod_{(e,o) \in {\cal E}} (1/X_{e,o})$. We then multiply by the vertex factors ${\cal L}^{(2m+2)}$ for every even-gon in ${\cal E}$.  

More crucially, the interpretation given to the vertex factors/ Feynman rules allows us to canonically write the amplitude as a sum over triangulations just like we do for a cubic theory. This is not the case for a random theory -- say including a Tr$(\Phi^6)$ coupling -- since there is no canonical way of ``blowing up'' a given 6-gon into a sum over its 14 internal triangulations; the most symmetrical choice would be to average over these 14 triangulations, with factors of $1/14$ appearing everywhere in the expressions, and in general for Tr$(\Phi^l)$ factors of $1/C_{l-2}$ for the average over the Catalan number of triangulations of the $l$-gon (as explained in~\cite{tropL}). But for the NLSM the presence of the Catalan factors in ${\cal L}^{(2m+2)}$ itself gives a completely canonical way of distributing these even-point contact interactions into triangulations. 

Consider a triangulation $\mathcal{T}$ of the $2n$-gon, containing $(2n-3)$-chords corresponding to the propagators, $X_{i,j} \in \mathcal{T}$. We start by identifying the chords in $\mathcal{T}$ that are \textbf{allowed} propagators of the NLSM, $i.e.$ all $X_{o,e}$. These chords divide the momentum $2n$-gon into different \textbf{even}-point sub-polygons, $\mathcal{P}_i$, for which we specify what we call the ``Catalan'' numerator factor
\begin{equation}
   {\cal N}_{{\cal T}} =  \prod_{\mathcal{P}_i}  \left(\sum_{X_{i,j}\in \mathcal{P}_i \cap \mathcal{T}} (- X_{i,j}) \right).
    \label{eq:reprule}
\end{equation} 

\begin{figure}[t]
    \centering
\includegraphics[width=0.4\textwidth]{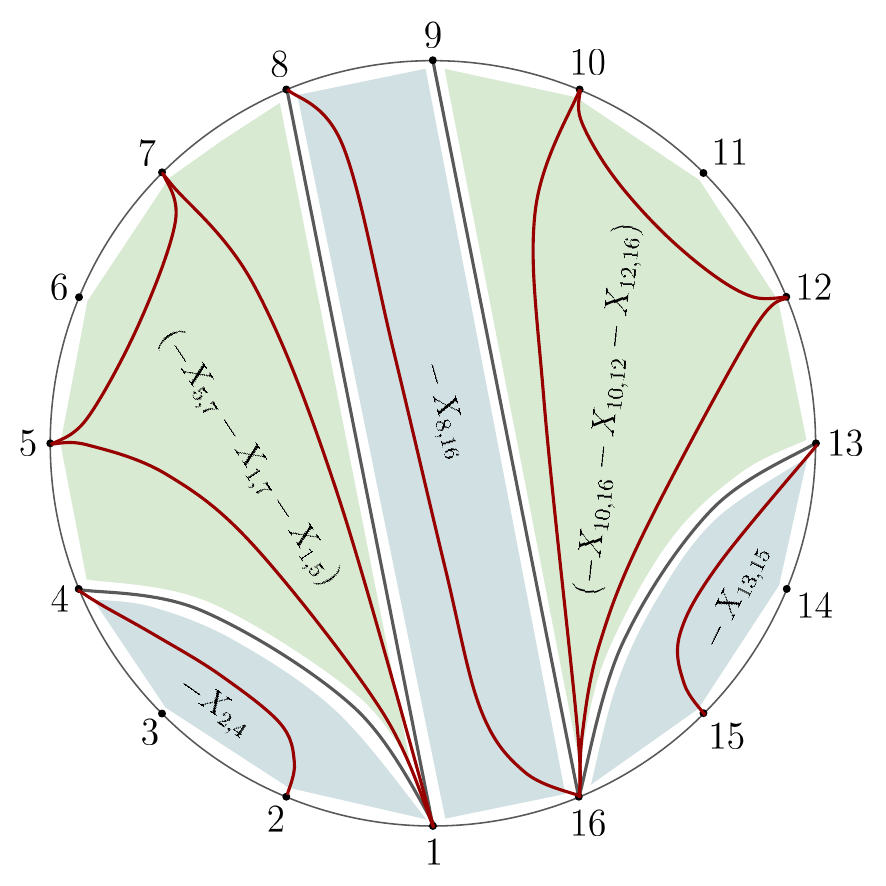}
    \caption{The Catalan numerator factor associated with a triangulation of the $2n=16$-gon. Each even-gon in the even-angulation is further triangulated by even-even or odd-odd chords. The even-gon then contributes a factor of the sum of the $(-X)'s$ for the interior chords that triangulate it.} 
    \label{fig:cayleycat}
\end{figure}

Said in words and illustrated in fig. \ref{fig:cayleycat}, given a triangulation ${\cal T}$, the $X_{e,o}$ chords divide the $2n$-gon in a collection of even-gons; we then take the sum of all the $(-X_{e,e}, -X_{o,o})$ inside each even-gon, and take the product of these sums over all the even-gons, to get the numerator factor ${\cal N}_{{\cal T}}$ associated with ${\cal T}$. We multiply this by the poles seen in ${\cal T}$, $\prod_{{e,o} \in \mathcal{T}} (1/X_{e,o})$ , and finally sum over all $\mathcal{T}$ to get the full NLSM amplitude. Note again that the chords entering the numerator are $X_{e,e}$ or $X_{o,o}$ as all the $X_{o,e}$ are bounding the $\mathcal{P}_i$'s. If there are \textbf{no} $X_{o,e}$, then $\mathcal{P}$ is the full $2n$-gon, and we get precisely the contact part of the amplitude. 

We can illustrate with the first interesting example of the $n=6$ amplitude, where we have a total of 14 triangulations. We have the triangulations where the $(e,o)$ chord $(1,4)$ splits the hexagon into two squares, with the $2 \times 2 = 4$ ways of triangulating each of the squares. These contribute with $(-X_{1,3} - X_{2,4})(-X_{1,5} - X_{4,6})/X_{1,4}$ to the amplitude. Taken together with the cyclic images of this for the $(e,o)$ chords being $(2,5)$, ($3,6)$, we have 12 of the 14 triangulations.  Finally, we have the contact term with no $(e,o)$ chords. There are 2 triangulations of these, the purely $(e,e)$ one with the chords $(2,4),(2,6),(4,6)$ contributing $(-X_{2,4} - X_{4,6} - X_{2,6})$ to the amplitude and the purely $(o,o)$ one with $(1,3),(3,5),(1,5)$, contributing $(-X_{1,3} - X_{3,5} - X_{1,5})$ to the amplitude, correctly giving the amplitude as 
\begin{equation}
\begin{aligned}
    {\cal A}_{6} = &\frac{(-X_{1,3} - X_{2,4})(-X_{1,5} - X_{4,6})}{X_{1,4}} + {\rm cyclic} \\
    &+ (-X_{2,4} - X_{4,6} - X_{2,6} - X_{1,3} - X_{3,5} - X_{1,5}).
\end{aligned}
\end{equation}

\section{Discovering shifted Tr($\Phi^3$)}

We are now very close to the representation in terms of shifted Tr$(\Phi^3)$~\cite{deltaNLSM}. We can first cheat by taking a look at the shifted answer. For any triangulation ${\cal T}$, we have $X_{e,o}$, $X_{e,e}$ and $X_{o,o}$ propagators, and the contribution from ${\cal T}$ to the shifted amplitude is 
\begin{equation}
\begin{aligned}
&\prod_{e,o} \frac{1}{X_{e,o}} \prod_{e,e} \frac{1}{X_{e,e} + \delta} \prod_{o,o} \frac{1}{X_{o,o} - \delta}  =   \\
&=\frac{(-1)^{N_{o,o}}}{\delta^{N_{e,e} + N_{o,o}}}\prod_{e,o} \frac{1}{X_{e,o}} 
 \prod_{e,e} \left(1 - \frac{X_{e,e}}{\delta} + \frac{X_{e,e}^2}{\delta^2} - \cdots \right)  \prod_{o,o} \left(1 + \frac{X_{o,o}}{\delta} + \frac{X_{o,o}^2}{\delta^2} + \cdots \right) ,\
\end{aligned}
\end{equation}
where in the second line we Taylor-expanded in $1/\delta$. Let's choose units in which $\delta=1$, then all the terms from the second line can be simply interpreted, and we can summarize the contribution of any ${\cal T}$ as follows: in addition to the product over all $1/X_{e,o}$ poles, we have a sum over {\it all} monomials in $(-X_{e,e})$ and $(+X_{o,o})$, with unit coefficient, and an overall sign given by the parity of the total number of $(o,o)$ chords in ${\cal T}$, $(-1)^{N_{o,o}}$. 

Note this already has similar ingredients to the Catalan representation of the NLSM. The difference is that apart from the peculiar signs, this shifted Tr$(\Phi^3)$ result is a completely democratic sum over all the $X_{e,e},X_{o,o}$ chords, whereas the Catalan form picks single $X$'s from the interior of each even-gon in ${\cal T}$. 

Nonetheless, we now show that the sum over all triangulations of the two forms exactly agrees. Let us consider all triangulations that have a fixed set of $(e,o)$ chords, differing only in their $(e,e)$ and $(o,o)$ ones. The set of $(e,o)$ chords defines an even-angulation ${\cal E}$ of the $2n$-gon. Each $2m$-gon of the even-angulation is further triangulated, by purely $(e,e)$, or $(o,o)$ chords. 
\begin{figure}[t]
    \centering
\includegraphics[width=0.4\textwidth]{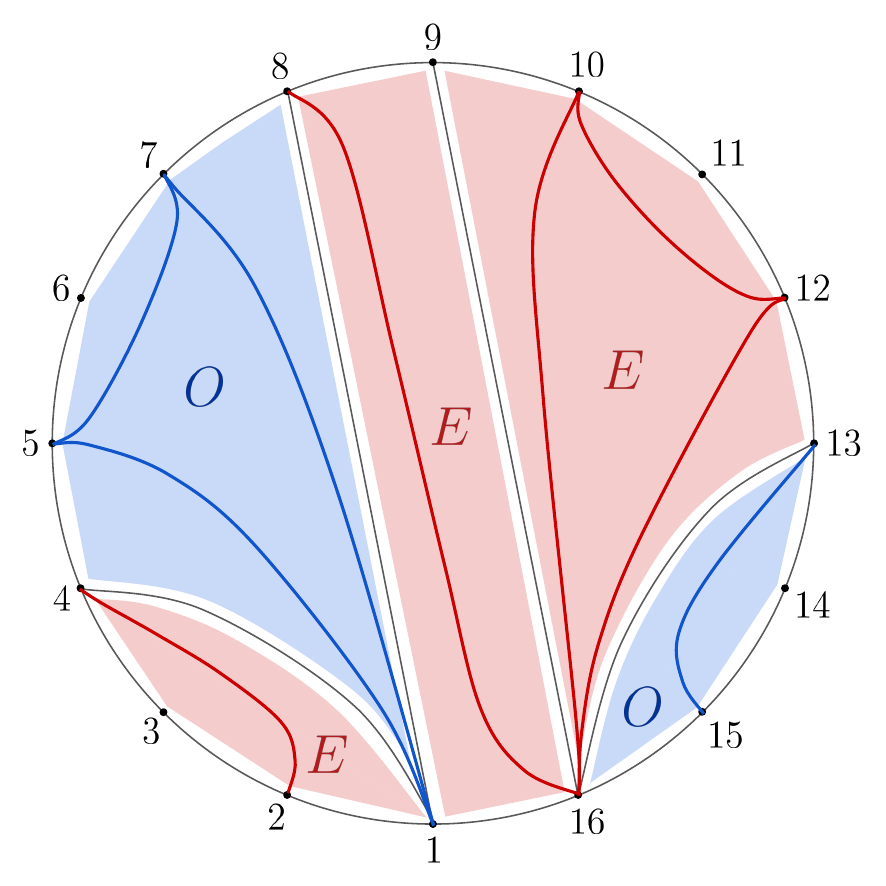}
    \caption{Any even-gon in an even-angulation is further triangulated either purely by $(e,e)$ or $(o,o)$ chords, and hence can be assigned an ``even'' or ``odd'' parity, as illustrated here for our running $2n = 16$ example.}
    \label{fig:EvenOdd}
\end{figure}
Therefore, we can assign each $2m$-gon of the even-angulation a ``parity'' of even or odd, respectively (see fig. \ref{fig:EvenOdd}). Importantly, there is a natural bijection between the ``even-even'' and ``odd-odd'' triangulations: given any ``even-even'' triangulation, we can simply shift all the indices by one to the right to get an ``odd-odd'' triangulation. This natural pairing of even-even and odd-odd triangulations will be useful in what follows. 

From the sum of all triangulations sharing the same ${\cal E}$, exactly one set of terms matches the Catalan representation: the one where the monomial in $(-X_{e,e}), (+X_{o,o})$ is obtained by picking a single chord from each $2m$-gon in ${\cal E}$. Note that the factor $(-1)^{N_{o,o}}$ is the  same for all the ``odd'' $2m$-gons in ${\cal E}$, since $(-1)^{N_{o,o}} =  (-1)^{2m - 3} = (-1)$. Therefore, each ``odd'' $2m$-gon picks a minus sign, and so the final monomial becomes precisely the product of $(-X)$ picking a single chord from the interior of each $2m$-gon -- precisely the Catalan rule. Note that this monomial has overall weight given by the total number of $2m$-gons in ${\cal E}$, $(N_{e,o} + 1)$.  We are left to show that all other monomials of weight $\leq (N_{o,e} + 1)$ cancel in the sum over all the triangulations. Let's start by considering the monomials of weight $(N_{e,o} + 1)$, but which acquire this weight by picking more than one $X$ from any $2m$-gon in ${\cal E}$. This means there must be at least one $2m$-gon, ${\cal P}$, in ${\cal E}$ from whom no $X$ is taken. But then there is another triangulation producing exactly the same monomial but with opposite sign: since the ``even-even'' and ``odd-odd'' triangulations are naturally partnered in pairs, we simply flip the parity of ${\cal P}$ to that of its partner, and these monomials cancel in pairs. The same argument shows that for any monomial with weight less than $(N_{e,o} +1)$, where by necessity some ${\cal P}$ can not have any $X$'s chosen, we have pairwise cancellation of these monomials in the sum. Note that, restoring the $\delta$ dependence, the power in $\delta$ of this leading term is $\delta^{-(N_{e,e} + N_{o,o})} \times \delta^{-(N_{o,e} + 1)} = \delta^{2n - 2}$, and we have seen that all the lower terms in the $\delta$ expansion cancel.  

We have thus established that the leading term in the $\delta$ expansion of the shifted Tr$(\Phi^3)$ amplitude agrees with the Catalan form of the NLSM. Note that the most peculiar feature of the Tr$(\Phi^3)$ kinematic shift -- the relative minus sign between the $(e,e)$ and $(o,o)$ chords, is crucial to match the Catalan form, owing to the fact that even-gons can be assigned an even/odd parity according to their internal triangulations.

\section{Loops}
We can easily extend this analysis to the loop integrand for the NLSM; we will describe this in detail at 1-loop and indicate the extension to all loop orders. 

The kinematic variables at one loop are associated with a $2n$-gon with a single puncture $p$. In addition to the usual $X_{i,j}$ kinematic variables for the external momenta, we have variables $X_{p,j}$ touching the puncture, encoding the loop momenta. 

In order to obtain the integrand associated with the minimal Lagrangian, we simply consider an even-angulation ${\cal E}$ of the punctured disk, where the internal puncture must occur on the boundary of some even-gon. 
\begin{figure}[t]
    \centering
\includegraphics[width=\textwidth]{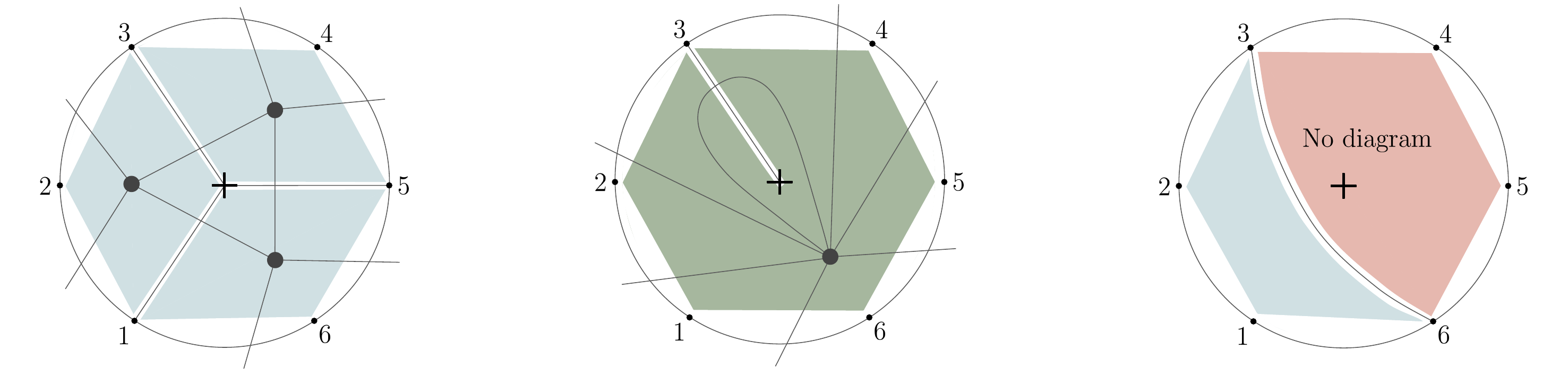}
    \caption{Having assigned a parity to the puncture (even in this example), the collection of $(e,o)$ chords defines an even-angulation of the punctured disk. Those even-angulations for which the $(e,o)$ chords touch the puncture are dual to Feynman diagrams. But those for which the puncture is contained inside an even-gon don't correspond to any diagram}. 
    \label{fig:looprule}
\end{figure}
We then associated the factor ${\cal L}^{(2m + 2)}$ with every $(2m+2)$-gon in ${\cal E}$, and add the poles associated with the edges of the even-gons. Note that at tree-level we could easily characterize both even-angulations as well as the allowed poles as associated with the $(e,o)$ chords. We can do exactly the same thing at one-loop, except that in defining what we mean by ``even/odd'' we have to assign the puncture a parity, and consider both possibilities of ``even/odd'' assignments. 

Now, we can translate this rule into a sum over triangulations in exactly the same way as we did at tree-level. We assign the puncture a parity, and look at any triangulation ${\cal T}$ of the punctured disk. The set of $(e,o)$ chords define an even-angulation ${\cal E}$. Now, consider those triangulations for which the $(e,o)$ chords in the associated ${\cal E}$ touch the puncture, i.e. for which the puncture is included in the boundary of at least one of the even-gons. These correspond exactly to the diagrams coming from the minimal Lagrangian. If we include exactly the same Catalan numerator factor ${\cal N}_{{\cal T}}$ defined above, then the sum over all the ${\cal T}$
with this same ${\cal E}$ reproduces ${\cal L}^{(2m+2)}$ for each even-gon in ${\cal E}$ just as at tree-level, matching the integrand we get from the minimal Lagrangian. 

But we are not quite done yet, because there are other triangulations for which the set of $(e,o)$ chords do not touch the puncture, or what is the same, where the puncture is in the interior of an even-gon. These never occur as diagrams coming from the minimal Lagrangian, and so we must give triangulations of this form vanishing weight in order to match the one-loop integrand from the Feynman rules of the minimal Lagrangian, see fig. \ref{fig:looprule} for examples of triangulations that do and don't correspond to diagrams from the minimal Lagrangian.  

The connection with the shifted Tr$(\Phi^3)$ theory then proceeds exactly as at tree-level for those triangulations where the $(e,o)$ chords touch the puncture. Every even-gon either contains no puncture, or has the puncture on the boundary. In both cases exactly as at tree-level, each even-gon can be assigned to be ``even'' or ``odd'' depending on whether they are further triangulated by $(e,e)$  or $(o,o)$ chords, with the same bijection between these assignments. Precisely the same cancellations as seen at tree-level occur in the sum over all these triangulations, and this matches the one-loop integrand from the minimal Lagrangian. 

But we also have those triangulations for which the even-gons contain the puncture in their interior. There will be one even-gon $E_p$ in ${\cal E}$ that contains $p$. We now have to further triangulate the interior of $E_p$. Suppose the puncture is assigned to be odd. We know that in the triangulation, there must be some chord that touches the puncture, and it can't be $(e,p)$, since we are assuming no $(e,o)$ chords touch $p$. But this means that all the chords in the triangulation must involve purely odd vertices. 
\begin{figure}[t]
    \centering
\includegraphics[width=0.3\textwidth]{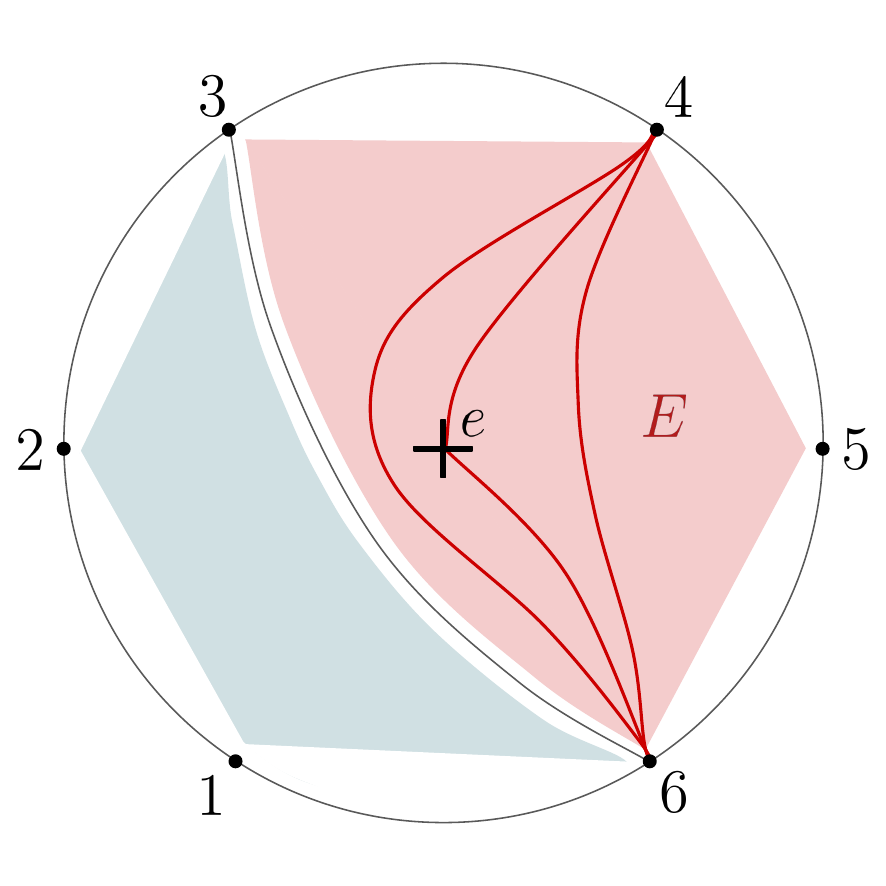}
    \caption{Even-gons that contain a puncture can only be triangulated with chords that have the same parity as the puncture. In this example the puncture is even, and the 4-gon containing the puncture can only be triangulated with $(e,e,)$ chords.}
    \label{fig:pureEven}
\end{figure}
Thus unlike the case of even-gons with no puncture in the interior, which can have one of two parities with a bijection between them, if the even-gon contains a puncture, it can have only one parity, the same as that of the puncture, as illustrated in fig. \ref{fig:pureEven}. In summing over triangulations that share this fixed ${\cal E}$, we will get the same cancellations at tree-level for all the other even-gons, but we simply get a factor of $+1$ from the leading term in the $\delta$ expansion from the chords in $E_p$. Summing over the two parity assignments for the puncture gives us a factor of $(+2)$ for the $E_p$ containing a puncture. 

Thus after summing over all triangulations and parity assignments to the puncture, the shifted Tr$(\Phi^3)$ theory gives us exactly the same rules as the Catalan  representation from the Lagrangian for all the triangulations where the associated even-gons don't contain the puncture in their interiors. But for those triangulations where an even-gon does contain the puncture, the minimal Lagrangian assigns the even-gon $E_p$ containing the puncture the factor $0$, while shifted Tr$(\Phi^3)$ assigns it the factor $(+2)$. But note that very nicely, precisely in these triangulations, there are no poles associated with any of the loop variables! Hence the difference between the integrand obtained directly from the minimal Lagrangian, and that obtained from the shifted Tr$(\Phi^3)$ theory, are scale-less integrals that integrate to zero. They are therefore both perfectly good integrands that give the same integrated amplitudes. 

These ``+1'' factors appear quite peculiar at first -- what is their purpose in life? We will see in the next section that, with a natural notion of ``surface-soft'' limit, it is precisely this shifted Tr$(\Phi^3)$ integrand that gives us an exact Adler zero, directly at the level of loop integrands. 

\begin{figure}[t]
    \centering
    \includegraphics[width=\textwidth]{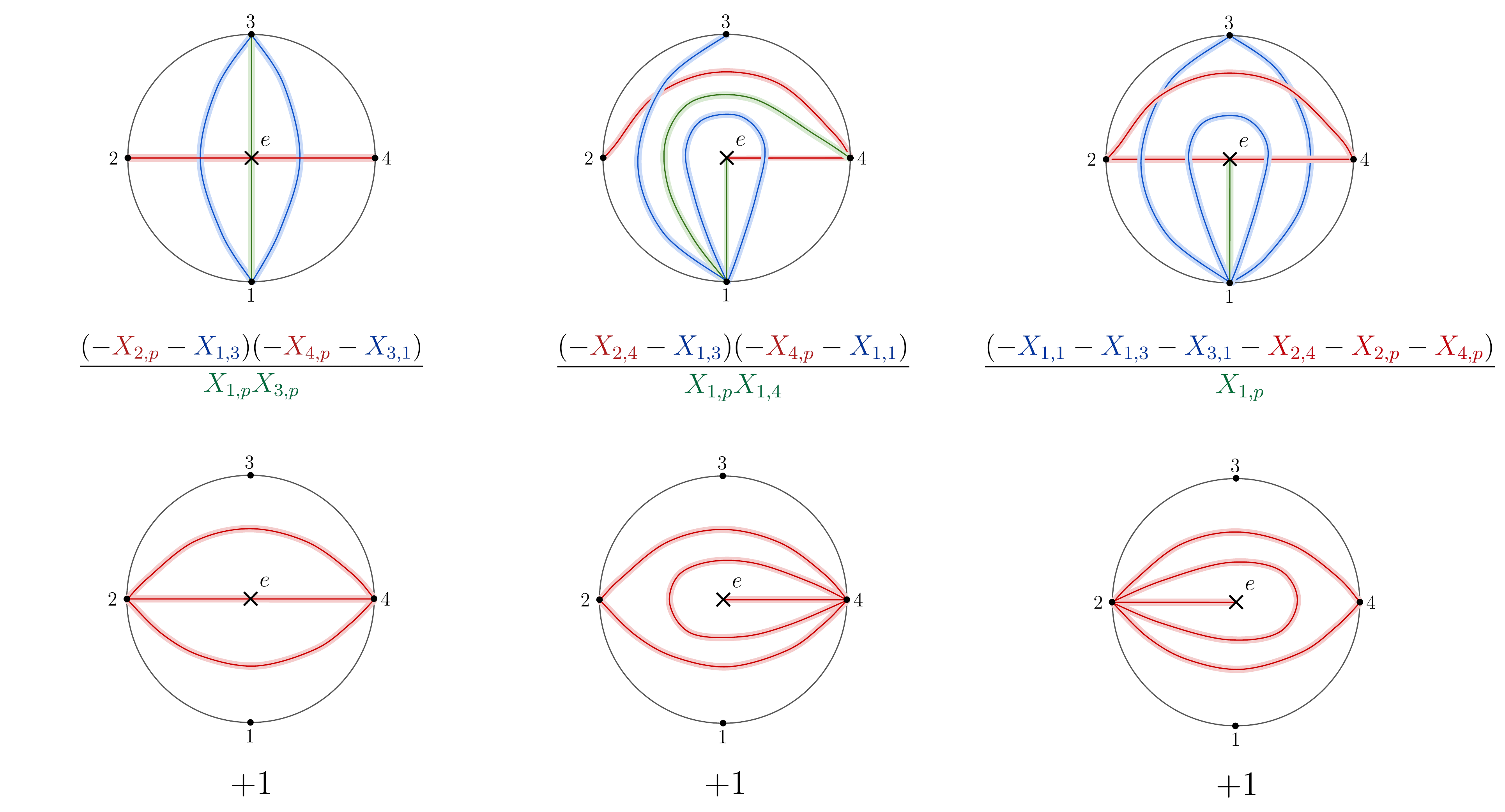}
    \caption{The one-loop rules. We assign a parity to the puncture. For every triangulation, we build the even-angulation using $(e,o)$ chords as usual. For those even-angulations where the $(e,o)$ chords touch the puncture, we have the usual Catalan rules, with a factor given by summing the $(-X)$ for the chords triangulating each even-gon. For those even-angulations with an even-gon $E_p$ containing the puncture in the interior, the minimal Lagrangian has vanishing weight, while the shifted Tr ($\Phi^3)$ assigns a factor $+1$ to $E_p$, which in this four-point examples happens to coincide with the full punctured disk.}  
    \label{fig:onelooprule}
\end{figure}
We illustrate these rules for the simplest case of $n=4$ pion scattering at one loop in fig. \ref{fig:onelooprule}, where all the different topologies of triangulations for the four-punctured disk are shown, together with the contributions associated with taking the even and odd chords in each even-angulation. We also include the triangulations where the puncture is contained inside an even-gon, which in this small example is simply the full disk. These are the contributions that are missing from the integrand coming directly from the minimal Lagrangian, but which are given a weight $+1$ from the shifted Tr$(\Phi^3)$ integrand. Putting the contributions from all the triangulations together gives us the $2n=4$ pion 1-loop integrand: 
\begin{equation}
\begin{aligned}
& {\cal I}^{{\rm 1-loop}}_{2n=4} = 6  + \frac{1}{X_{1,4}} \left[-2 X_{1,3} - 2 X_{2,4} + \frac{X_{1,1} X_{1,3} + X_{1,1} X_{2,4} + X_{1,3} Y_4 + X_{2,4} Y_4}{Y_1} \right.\\ 
& + \left. \frac{X_{1,3} X_{4,4} + X_{2,4} X_{4,4} + X_{1,3} Y_1 + X_{2,4} Y_1}{Y_4}  \right]  + {\rm cyclic} \\
& + \frac{-X_{1,1} - X_{1,3} - X_{2,4} - X_{3,1} - Y_2 - Y_4}{Y_1} + {\rm cyclic} \\ 
& + \frac{X_{1,3} X_{3,1} + X_{3,1} Y_2 + X_{1,3} Y_4 + Y_2 Y_4}{Y_1 Y_3} + \frac{X_{2,4} X_{4,2} + X_{2,4} Y_1 + X_{4,2} Y_3 + Y_1 Y_3}{Y_2 Y_4}.
\end{aligned}
\end{equation}

We finally sketch the story for general loop orders, considering for simplicity the planar limit where we have an $L$-punctured disk at $L$-loops. Consider all the $2^L$ even/odd assignments for the punctures, and the $(e,o)$ chords of any triangulation ${\cal T}$ define an even-angulation ${\cal E}$. The integrand from the minimal Lagrangian is associated with those even-angulations for which the punctures all occur on the boundaries of some even-gon in ${\cal E}$; all even-gons that contain punctures must be assigned vanishing weight. The shifted Tr $(\Phi^3)$ theory will agree with this integrand for all such triangulations, but will instead assign non-vanishing factors to the even-gons that contain punctures. However, these will differ from the integrand coming from the minimal Lagrangian by terms that integrate to zero. 

\section{Adler zero from Catalan}

We will now show that the representation of the NLSM tree amplitudes in the Catalan form, as a sum over triangulations, gives a very simple and direct understanding of the Adler zero for pion amplitudes as the pion momentum $p_{2n}^\mu \to 0$ is made soft. This is cleanly translated into a statement in terms of the planar variables $X_{i,j}$. In the soft limit, we have $X_{j,2n} \to X_{1,j}$. In particular, $X_{1, 2n-1} \to X_{2n-1,2n} = 0$ and $X_{2,2n} \to X_{1,2} = 0$. Note that these variables do not correspond to poles. 
We will generalize this to a notion of ``surface-soft'' limit at general loop orders in the next section. 

The argument for the Adler zero is naturally inductive, beginning with the obvious fact that the four-particle amplitude vanishes in the soft limit. We show that if the $2(n-1)$ point amplitude vanishes in the soft limit, the $2n$ point amplitude also vanishes. 

To begin with, note that the Catalan rules make it manifest that the amplitude, ${\cal A}$, is at most linear in $1/X_o$ for the chords $(i,j)$ with $|i-j|$ odd and $X_e$ for the ones where $|i-j|$ is even. Let's choose an odd chord $X_o$ which does not touch either index $1$ or $2n$. We can write the amplitude as follows:
\begin{equation}
    {\cal A} = \frac{\alpha}{X_o}  + \beta,
\end{equation}
where $\alpha, \beta$ don't depend on $X_o$. Now we know that as $X_o \to 0$, by factorization we must have that $\alpha = {\cal A}_L \times {\cal A}_R$, $i.e.$ the amplitude factorizes into lower-point NLSM amplitudes. But quite nicely, since $\alpha$ doesn't depend on $X_o$, then we have that $\alpha = {\cal A}_L \times {\cal A}_R$ even away from the locus setting $X_o \to 0$. The soft particle always occurs in one of the lower point factors and so, by the inductive assumption, we have that $\alpha$ must vanish in the soft limit. Since we can make this argument for any $X_o$ not touching $1$ or $2n$, we learn that the only terms in the amplitude that survive in the soft limit can \textbf{only} have poles of the form $1/X_{1,2j}$, $1/X_{2k+1, 2n}$, or be a pure contact term with no poles at all. 

Let's now look at the possible dependence of the amplitude on one of these variables, say $X_{1,2j}$. Since we know that $X_{2j,2n} \to X_{1,2j}$ in the soft limit, it is important to look at the dependence on $X_{2n,2j}$ as well. So we consider the terms in the amplitude of the form:
\begin{equation}
    {\cal A}=\frac{(a X_{2j,2n} +b)}{X_{1,2j}}  + c,
\end{equation}
where $a,b$ don't depend on either of $X_{1,2j},X_{2j,2n}$. Now again consider the factorization channel where $X_{1,2j} \to 0$, then the combination $(a X_{2j,2n} + b) = {\cal A}_L \times {\cal A}_R$ by factorization. But now in the soft limit of ${\cal A}_R$, we have that $X_{2j,2n} \to 0$, and hence we can only conclude that $b \to 0$. Since we can make this argument for every possible pole $X_{1,2j},X_{2k+1,2n}$ that can touch $1$, $2n$, we learn that the only terms in the amplitude that can survive the soft limit that have any poles at all, can have a product of poles of the form $1/X_{1,2j}$ and/or $1/X_{2k,2n}$, but with a numerator that must include a factor $X_{2j,2n}$ for every $1/X_{1,2j}$ and $X_{1,2k}$ for every $1/X_{2k,2n}$. 

From the Catalan rule, we know that the numerator factors can only come from chords inside the even-gons. But it is trivial to see that this is impossible if there is more than one pole!  Once we arrange that the poles touching $1,2n$ are non-crossing, if there is more than one $(1,2j)$ or $(2k+1,2n)$ chords it is obvious that the even-gons can never contain the needed numerator factors are not contained inside the even-gons. We can then consider the case of a single $(1,2j)$ and $(2k+1, 2n)$ chord. But then the chords $(1, 2k+1)$ and $(2j, 2n)$ necessarily cross in the quadrilateral formed by $(1,2j,2k+1,2n)$, and hence we can't pick them both as needed in the numerator.  

We thus conclude that the only terms in the amplitude that can survive the soft limit are of the form 
\begin{equation}
\sum_j \frac{a_{2j} (-X_{2j,2n})}{X_{1,2j}} + \sum_k \frac{b_{2k + 1} (-X_{1,2k+1})}{X_{2k+1,2n}} + c,
\end{equation}
which, in the soft limit, becomes 
\begin{equation}
S=-\sum_j a_{2j} - \sum_k b_{2k+1} + c.
\end{equation}

Finally, it is easy to see that $S$ vanishes via cancellation by pairs. Note that the terms $a_{2 j} (-X_{2j,2n})/X_{1,2j}$ come from triangulations containing the single odd chord $X_{1,2j}$; we must take the factor $X_{2j,2n}$ from the right even-gon and hence the right even-gon must be of even parity. The $a_{2 j}$ can be taken from the left even-gon, which can be of either odd or even parity. Analogous statements can be made about the $b_{2k+1}$ terms. Meanwhile, the contact term $c$ comes from summing over triangulations of the full $(2n)$-gon, which can be either even or odd. 
\begin{figure}[t]
    \centering
\includegraphics[width=0.8\textwidth]{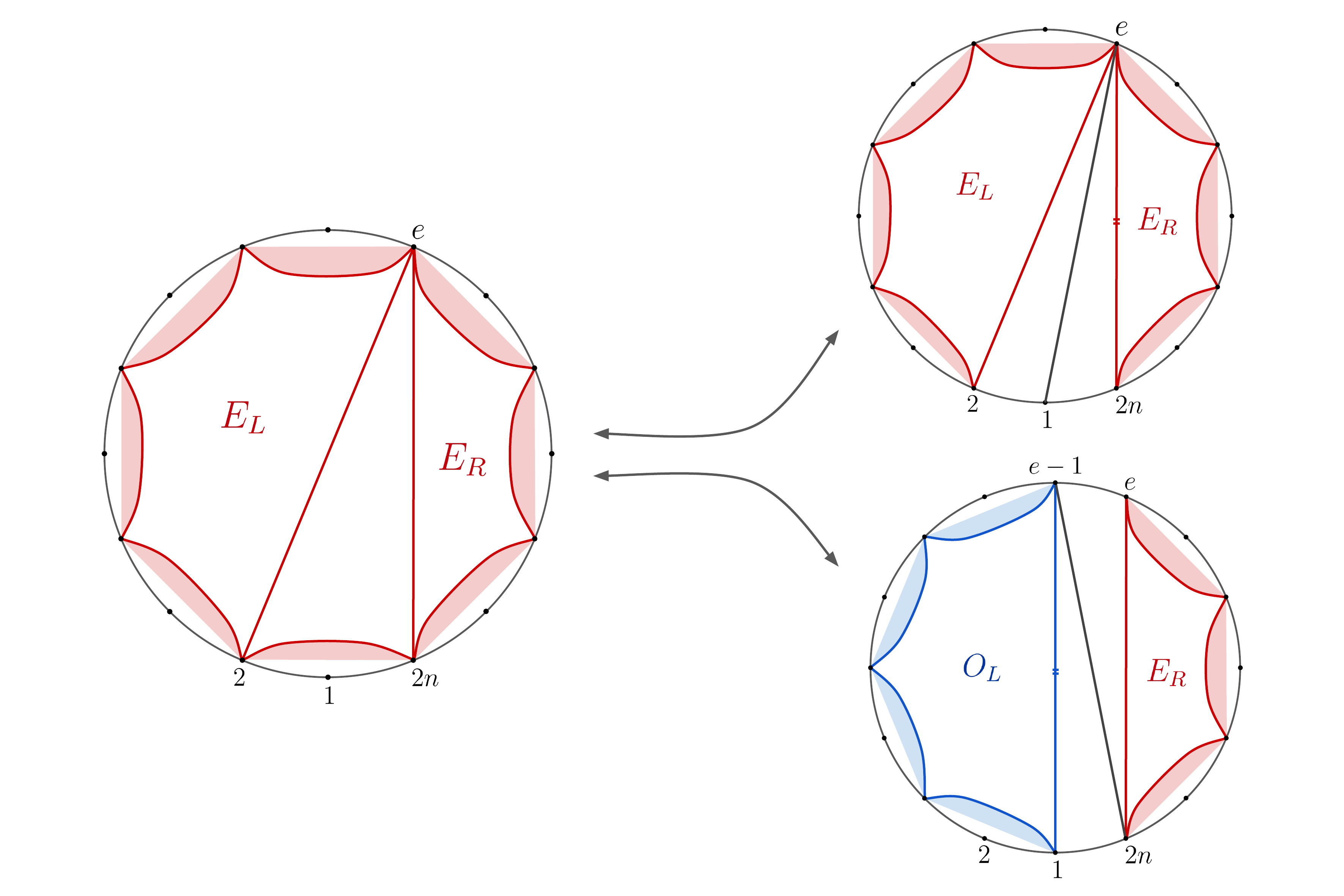}
    \caption{The natural bijection between triangulations contributing to the single-pole and contact-term contributions to the soft limit, canceling in pairs.}
    \label{fig:bijection}
\end{figure}

But there is a natural bijection between the triangulations contributing to $a_{2j},b_{2k+1}$ and the ones contributing to $c$, that allows us to see all the terms that can appear cancel in pairs. Consider for instance any $X_{e,e}$ variable. In $c$, it must occur in some purely $(e,e)$ triangulation ${\cal T}_e$ of the $2n$-gon. Now in any such ${\cal T}_e$, the chord $(2,2n)$ must appear as part of some triangle together with a  third even vertex $e$. But this allows us to naturally make a pairing between ${\cal T}_E$ and two different triangulations of the $(2n)$-gon. In the first, we simply delete the chord $(2,2n)$ and add back in $(1,e)$. In the second, we delete the chord $(2,2n)$ but now add back in $(e-1,2n)$. On the left side we need to have an $(o,o)$ triangulation, $O_L$, since we need to pick $X_{e-1,2n}$, but on the right, we pick an $(e,e)$ triangulation as seen in fig. \ref{fig:bijection}. Note that for $c$, we can get contributions to $c$ from $X_{e,e}$ from $E_L, E_R$ as well as $X_{2,2n}$, but this last term vanishes in the soft limit, so we only have the contributions from $X_{e,e}$ in $E_L,E_R$. But for every contribution from an $X_{e,e}$ in $E_L$ to $c$, we have a contribution to $a_{e}$ where $(-X_{2n,e})$ is chosen from $E_R$ and $X_{e,e}$ is chosen from $E_L$. And for every contribution to $c$ from $E_R$, we have a contribution to $b_{e-1}$ where $(-X_{e-1,2n})$ is chosen from $O_L$ and $X_{e,e}$ is chosen from $E_R$. In this way all the terms in the soft limit $S = (-\sum_j a_{2j} - \sum_k b_{2k + 1} + c)$ can be seen to cancel and in the soft limit, we get $S=0$, yielding the Adler zero as desired.

\section{Surface-soft limits and the Adler zero for the loop integrand}

The existence of the Adler zero follows from the non-linear realization of the chiral symmetry and holds for the full integrated amplitude at all loop orders. But it is interesting to ask about the behavior of the loop integrand in the soft pion limit, at least in the planar limit where the notion of ``loop integrand'' is well-defined. This question has been the subject of very interesting recent investigations \cite{JaraSoftTheorem,LoopRecursionJara}, with the conclusion that the Adler zero is {\it not} present at the level of the loop integrand. Instead the soft limit yields integrands that are associated with scale-less integrals that integrate to zero. Of course, at loop level there are multiple notions of ``loop integrand'' that can come from different Lagrangians, differing by field redefinitions, or even more general modifications, which differ by scale-less integrals, integrating to the same final result. So it is interesting to ask whether there is a ``best possible'' integrand in the sense of having the nicest possible behavior in the soft limit, and \cite{JaraSoftTheorem} argued that this ``nicest possible'' integrand is precisely the one associated with the shifted Tr$(\Phi^3)$ theory. 

The reasons for subtleties involving the soft limit at the level of the loop integrand are quite clear and familiar. If we wish to deal with the integrand for on-shell amplitudes, we have the ubiquitous ``1/0'' difficulties associated with loops decorating the massless external lines. Thus some care must be taken to define what is even meant by soft limits to begin with. For instance at one-loop, \cite{JaraSoftTheorem} made the natural choice to allow the variables $``X_{i+2, i}$'' -- which are exactly the propagators associated with massless external bubbles -- to be non-zero, defining a sort of off-shell extension of the amplitude. 

In this section, we revisit this fundamental question about the Adler zero at the level of the loop integrand, but we will take a different point of view. Instead of worrying about defining the soft limit in terms of the external particle momenta, we will use the most natural notion of ``soft limit'' provided to us directly by surfaces. This will define an identification between curves on surfaces that also gives a mapping on the kinematic $X$ variables we will call the ``surface-soft'' limit. Of course, this new notion of soft limit agrees with the usual one for most kinematic invariants but resolves all the subtleties associated with bubbles and tadpoles. Quite beautifully with this definition of ``surface-soft'' limit, we will find a clean Adler zero directly at the level of the loop integrand! 
\begin{figure}[t]
    \centering
    \includegraphics[width=\textwidth]{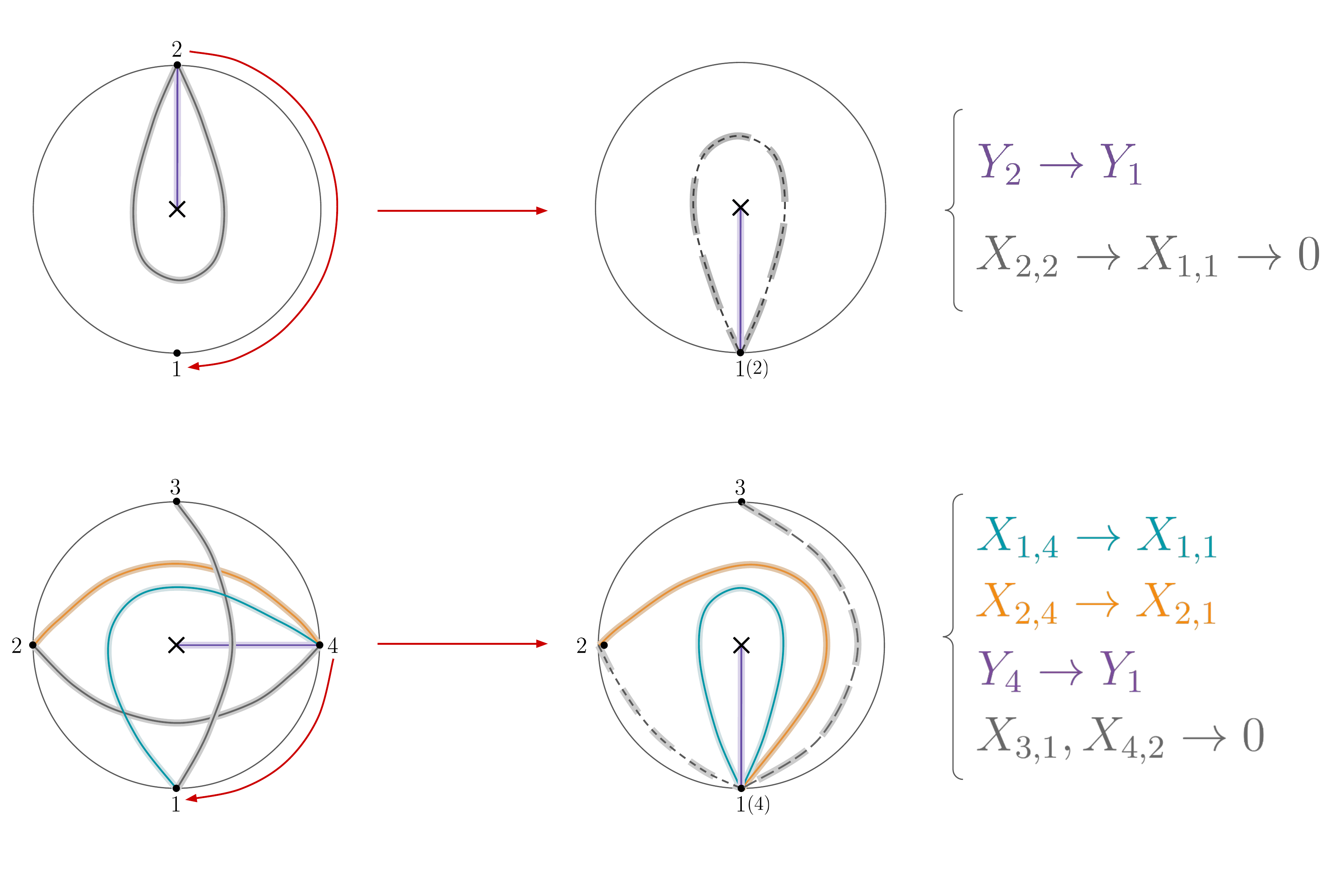}
    \caption{Examples of the mapping of curves and kinematics for the ``surface soft'' limit, at one-loop for two and four points, where the points 2 and 4 are put on top of 1 respectively.  The curves that become identified with boundaries are represented with dashed lines.}
    \label{fig:surfacesoft}
\end{figure}

The definition of ``surface-soft'' limit is straightforward. Suppose we consider the $2n$ pion amplitude, and take the limit where the last pion is soft $p_{2n}^\mu \to 0$. In the picture of the momentum polygon, this means that we are putting the point $2n$ on top of the point $1$. This is also a very natural operation to perform at the level of the surface. We begin with a surface which has a boundary with $2n$ marked points $\{1, 2, \cdots, 2n\}$ and we simply identify the points $\{1,2n\}$ to get a smaller surface where the boundary has the $2n-1$ marked points $\{1, 2, \cdots, 2n-1\}$. When we do this, a curve $C$ on the original surface is mapped to another curve $\tilde{C}$ on the smaller surface. Most of the $\tilde{C}$ will be interior curves of the smaller surface, and so we just map the corresponding kinematical variables in the obvious way with $X_C \to X_{\tilde{C}}$. Sometimes the curve $\tilde{C}$ can be a boundary of the smaller surface. For these, we have the ``on-shell'' condition that $X_{\tilde C} = 0$, and so for those $C$'s that map to boundaries, we send $X_C \to 0$.

Note that this is an especially clean definition of the soft limit -- since we are taking the soft limit directly at the level of the kinematic invariants $X_C$ we are trivially preserving momentum conservation, in contrast with the more usual definition rescaling the momentum $p_{2n}^\mu \to t \, p_{2n}^\mu$, with $t\to 0$, where we implicitly assume the remaining momenta are changed to preserve momentum conservation. 
 
We have already seen how this works at tree-level. When we put the point $2n$ on top of $1$, all the curves $(j,2n)$ become identified with $(1,j)$. For $j=3,\cdots, 2n-2$, these are all interior curves of the smaller surface, and so we send $X_{j,2n} \to X_{1,j}$ for $j=3,\cdots, 2n-2$. But for $j=2,2n-1$, the curves $(1,j)$ become boundaries, and so we send $X_{2,2n}, X_{1,2n-1} \to 0$. See fig. \ref{fig:surfacesoft} for examples at 1-loop for 2 and 4 points, which we will now work through in more detail. 

Let us begin with the smallest example of the $n=2$ amplitude at one-loop.  This case may appear quite degenerate, but as we'll see everything will end up being perfectly well-defined. Let's begin with the Tr $(\Phi^3)$ integrand for the bubble, which is simply 
\begin{equation}
    {\cal I}_{2n=2} = \frac{1}{Y_1 Y_2} + \frac{1}{Y_1 X_{1,1}} + \frac{1}{Y_2 X_{2,2}},
\end{equation}
here $Y_{1,2}$ are the loop variables going from $(1,2)$ to the puncture, and $X_{1,1}, X_{2,2}$ are the two tadpole variables. Now to get the NLSM integrand, we assign the puncture a parity, and perform the $(e,e), (o,o)$ shifts. Assigning the punctures to be even/odd gives us 
\begin{eqnarray}
    {\cal I}^{{\rm even}}_{2n=2, \delta} &=& \frac{1}{Y_1 (Y_2 + \delta)} + \frac{1}{Y_1 (X_{1,1} - \delta)} + \frac{1}{(Y_2 + \delta) (X_{2,2} + \delta)}, \nonumber \\ {\cal I}^{{\rm odd}}_{2n=2, \delta} &=& \frac{1}{(Y_1 - \delta) Y_2} + \frac{1}{(Y_1 - \delta) (X_{1,1} - \delta)} + \frac{1}{Y_2 (X_{2,2} + \delta)}.
\end{eqnarray}

To get the NLSM integrand, we add what we get from the two parity assignments ${\cal I}_{2n=2,\delta} = {\cal I}^{{\rm even}}_{2n=2, \delta} + {\cal I}^{{\rm odd}}_{n=2, \delta}$ and take the leading term in the $1/\delta$ expansion.  Now let us examine the surface-soft limit. When we put $2$ on top of $1$, the curves $(1,p)$ and $(2,p)$ (where $p$ is the puncture) become identified, and so we must have $Y_2 \to Y_1$. We also have that $(2,2)$ and $(1,1)$ become identified. But in this very small example, it also so happens that the curve $(1,1)$ on the smaller surface is the same as the boundary, and so we must send $X_{2,2}, X_{1,1} \to 0$. It is then simple to see that ${\cal I}^{{\rm even, odd}}_{2n=2, \delta} \to 0$ in this limit, for any finite $\delta$, as e.g.
\begin{equation}
{\cal I}^{{\rm odd}}_{2n=2, \delta} \to \frac{1}{\delta Y_1} + \frac{1}{\delta (Y_1 - \delta)} + \frac{1}{Y_1 (Y_1 - \delta)} = 0.
\end{equation}

Note that rather remarkably we don't need to add the two parities to get the Adler zero, nor did we need to take the leading term in the $1/\delta$ expansion to get the ``pure'' NLSM amplitude; we have a zero in the surface-soft limit for the surface-soft limit for each parity assignment separately, and for any finite $\delta$.

Let's also work through the next example with $2n=4$, where we encounter the behavior seen more generically for all larger $n$. When we put the point $4$ on top of $1$, the curve $(4,p)$ becomes identified with $(1,p)$ so we send $Y_4 \to Y_1$. We also have the curves $(4,j)$ which are identified with $(1,j)$, for $j=2,3$; (note that $(4,1)$ is a boundary curve on the big surface and isn't included). For $j=2$ we $(1,2)$ is a boundary so we must send $X_{4,2} \to 0$. Similarly $(j,4)$ is identified with $(j,1)$ for $j=2,3$. Since $(3,4)$ is a boundary we must send $X_{3,1} \to 0$. Finally the curves $(1,4),(4,4)$ are identified with $(1,1)$. In summary, the surface-soft limit for $2n=4$ consists of sending $Y_4 \to Y_1, X_{4,2} \to 0, X_{4,3} \to X_{1,3}, X_{3,1} \to 0, X_{2,4} \to X_{2,1}, X_{1,4} \to X_{1,1},X_{4,4} \to X_{1,1}$. 

We can now examine the integrand obtained from the shifted Tr $(\Phi^3)$ theory. Before shifts, the 4-pt 1-loop integrand is sum over 35 terms
\begin{equation}
\begin{aligned}
    & {\cal I}_{{\rm Tr}(\Phi^3), 4} = \frac{1}{Y_1 Y_2 Y_3 Y_4} + 
     \left[\frac{1}{Y_1 Y_2 Y_3 X_{3,1}} + {\rm cyclic} \right] + \frac{1}{Y_1 Y_3 X_{1,3} X_{3,1}} + \frac{1}{Y_2 Y_4 X_{2,4} X_{4,2}} \\ 
    & + \left[\frac{1}{Y_1 Y_2 X_{2,1}}\left(\frac{1}{X_{2,4}} + \frac{1}{X_{3,1}} \right) + {\rm cyclic} \right] \\
    & + \left[\frac{1}{Y_1 X_{1,1}} \left(\frac{1}{X_{1,3} X_{1,4}} + \frac{1}{X_{1,4} X_{2,4}} + \frac{1}{X_{2,4} X_{2,1}} + \frac{1}{X_{2,1} X_{3,1}} + \frac{1}{X_{3,1} X_{1,3}} \right) + {\rm cyclic} \right].
    \end{aligned}
\end{equation}

Now let's pick the puncture to have even parity. We then shift the kinematics as $Y_2 \to Y_2 + \delta, Y_4 \to Y_4 + \delta$ and $X_{e,e} \to X_{e,e} + \delta, X_{o,o} \to X_{o,o} - \delta$. Finally as we take the surface-soft limit as just defined, sending $Y_4 \to Y_1, X_{4,2} \to 0, X_{4,3} \to X_{1,3}, X_{3,1} \to 0, X_{2,4} \to X_{2,1}, X_{1,4} \to X_{1,1},X_{4,4} \to X_{1,1}$, the shifted integrand vanishes. 

Note that while for the conventional connection between the $X_{i,j}$ and momenta we would have $X_{i,j} = X_{j,i}$, this identification is not natural on the surface and we do not make it. So e.g. while $(1,2)$ is a boundary, $(2,1)$ is not and hence we do not set $X_{2,1}$ to zero. Similarly unlike the case of the bubble for $2n=2$, $(1,1)$ is not a boundary so we don't set any of the tadpole $X_{j,j}$ to zero. As we have seen allowing $X_{i,j} \neq X_{j,i}$ is crucial to obtaining the Adler zero for the integrand. A natural physical picture for this difference \cite{upcoming} is to imagine that the puncture itself carries a small momentum $q^\mu$; this would add the momentum $q^\mu$ to all curves surrounding the puncture. We can think of this as corresponding to an amplitude where in the addition to the colored pions we emit an uncolored scalar ``dilaton'' corresponding to the puncture. This amplitude is proportional to that of the pions as the dilaton momentum is taken to zero. But for the Adler zero, it is crucial to take the pion momenta to be even softer than the dilaton momentum, which regulates the ambiguities in the zero pion momentum limit. 

Having seen that we expect the Adler zero at the level of the integrand when taking the natural ``surface-soft'' limit, let's turn to proving that this happens at one-loop directly from the same sort of combinatorial arguments we saw with the proof of the Adler zero at tree-level. Note that this is already a more restricted statement than what we have observed to be true--the zero is present for any finite $\delta$, not just for the leading term in the $1/\delta$ expansion that gives us the NLSM. Nonetheless, our simple triangulation-rules for the NLSM are precisely for this leading term in the $1/\delta$ expansion, and for these, we will see a simple understanding of the zero paralleling what we observed at tree-level.

We have already explained how we build the NLSM loop integrand as a sum over triangulations of the punctured disk with $2n$-marked points on the boundary -- the surface describing the 1-loop $2n$-point amplitude. This rule requires assigning a parity to the puncture; to get the full NLSM loop integrand we needed to sum over both parity assignments. We will now see how the Adler zero for the one-loop integrand follows from these rules. We will in fact prove the stronger statement already seen in our examples, that the result we get from each parity assignment of the puncture vanishes in the soft limit individually. 

Once more the argument will be inductive. We have already checked explicitly that the $2n = 2$ point integrand for any parity assignment of the puncture vanishes in the soft limit.
We then fix a parity of the puncture and assume that the corresponding 1-loop $2(n-1)$ NLSM integrand vanishes in the soft limit, and we use this to prove that 1-loop integrand vanishes in the soft limit for $2n$ points. 

Just like at tree-level, the   Catalan rule at 1-loop tells us that the integrand should be linear in $1/X_o$ and $X_e$, where here $X_o$ denotes all curves in which the parity of the starting and end points is different, and  $X_e$ those with equal parity. So, for example, if the parity of the puncture is even, then $Y_{2k+1}$ are $X_o$ while $Y_{2k}$ are  $X_e$. 

Let's fix the parity of the puncture to be even. In addition, to make the notation uniform we denote the puncture variables, $Y_i = X_{0,i}$, so the index $0$ is the puncture label. Just like in the tree-level case, we can write the dependence of the integrand in any $X_o$ not touching indices $1$ or $2n$ as $\mathcal{I}^e_{2n} = \frac{\alpha}{X_o} + \beta$, 
where $\alpha$ and $\beta$ are independent of $X_o$. Once more by factorization in $X_o \to 0$, we have $\alpha=0$ in the soft limit. Therefore, just like in the tree once more we can focus on the dependence of $X_{1,2j}$ (or, equivalently in $X_{2k+1,2n}$):
\begin{equation}
    \mathcal{I}^e_{2n} = \frac{a_{0,1} (-Y_{2n}) +b_0}{Y_1} + \sum_{j=1}^n \frac{(a_{1,2j} (-X_{2n,2j})+b_{1,2j})}{X_{1,2j}} + \sum_{j=1}^n \frac{(a_{2j,1} (-X_{2j,2n})+b_{2j,1})}{X_{2j,1}} + c_1.
\end{equation}

Just like at tree-level, factorization on $X_{1,2j}=0$, $X_{2j,1} = 0$ allows us to conclude that $b_{1,2j}=0,b_{2j,1} = 0$ in the soft limit. The analogous statement holds for $X_{2k+1,2n}, X_{2k+1,2n}$ poles. So we conclude that, other than the pure contact part, the non-vanishing pieces in the soft limit can only have poles corresponding to chords touching $1$ or $2n$, and must be possible products of $X_{2n,2j}/X_{1,2j}$, $X_{2j,2n}/X_{2j,1}$, $X_{2k+1,1}/X_{2k+1,2n}$ and $X_{1,2k+1}/X_{2n,2k+1}$ that $\to 1$ in the soft limit. 

Now we know that the numerator factors correspond to chords inside even-gons, except if the even-gon contains the puncture, in which case we just get a factor of $(+1)$. At tree-level we were able to exclude the presence of products of more than one of such terms simply because it was impossible to consistently obtain the correct numerator factors for an even-angulation including more than one of the $X_o$. The same argument holds for any two $X_o$ that do not end on the puncture, as well as for collections of $X_o$ in which all end on the puncture. But we now need to understand what happens when we have both $X_o$ that end on the puncture and some that do not. 

It turns out that, just like in the tree case, we can have only one $(e,o)$ pole that touches $1$ or $2n$. The presence of the puncture doesn't affect the argument, because we can't have any triangles of the triangulation containing the puncture. Therefore, we conclude that in the soft limit the answer reduces to:
\begin{equation}
\begin{aligned}
    &\left( \sum_{j=1}^n \frac{\alpha_{1,2j} (-X_{2n,2j})}{X_{1,2j}}  +\frac{\alpha_{2j,1} (-X_{2j,2n})}{X_{2j,1}}  +\sum_{k=0}^{n-1}  \frac{\beta_{2k+1,2n} (-X_{2k+1,1})}{X_{2k+1,2n}} + \frac{\beta_{2n,2k+1} (-X_{1,2k+1})}{X_{2n,2k+1}} 
    \right. \\
    &\left. \quad + \frac{\alpha_{0,1} (-Y_{2n})}{Y_1} 
    + \gamma \right) \to -\alpha_{0,1} +  \sum_j (-\alpha_{1,2j} - \alpha_{2j,1}) + \sum_k (-\beta_{2n,2k+1} - \beta_{2k+1,2n}) + \gamma 
    \end{aligned}
\end{equation}
where $\alpha$ and $\beta$ don't have any poles for the reason explained and $\gamma$ is the pure contact part. We now want to show that, just like in the tree-level case, there is a pairwise cancellation between the contact part and $\alpha$, $\beta$ from which we derive the Adler zero of  $\mathcal{I}^e_{2n}$. The contact part comes from full even/odd-angulations. However, since we have fixed the parity of the puncture to be even, the contributions to $\gamma$ come only from pure $X_{e,e}$ triangulations, which each contribute with $+1$. 

Inside the pure $X_{e,e}$ triangulation contributing to the contact term, the edge $X_{2n,2}$ has to be contained in one of three types of triangles: 
\begin{enumerate}[label=\Roman{*}.]
    \item $\{X_{2j,2},X_{2j,2n},X_{2n,2}\}$, where the puncture is on the left;
    \item   $\{X_{2,2j},X_{2n,2j},X_{2n,2}\}$, where the puncture is on the right;
    \item $\{X_{0,2},X_{0,2n},X_{2n,2}\}$, where the puncture is a vertex of the triangle.
\end{enumerate}

Note we can never have a case where the triangle contains a puncture since this must always be further triangulated with internal chords. 

The number of pure $X_{e,e}$ triangulation containing this triangle is then the product of the number of $X_{e,e}$ triangulations of the left problem, $N_{e,e}^L$, times those of the right problem, $N_{e,e}^R$. So from this triangle, we get a contribution to the contact term of $N_{e,e}^L\times N_{e,e}^R$. If the triangle touches the puncture, then the contact is simply the number of triangulations of the lower-point tree problem,  $N_{e,e}^{\text{tree}}$. 

\begin{figure}[t]
    \centering
    \includegraphics[width=0.6\textwidth]{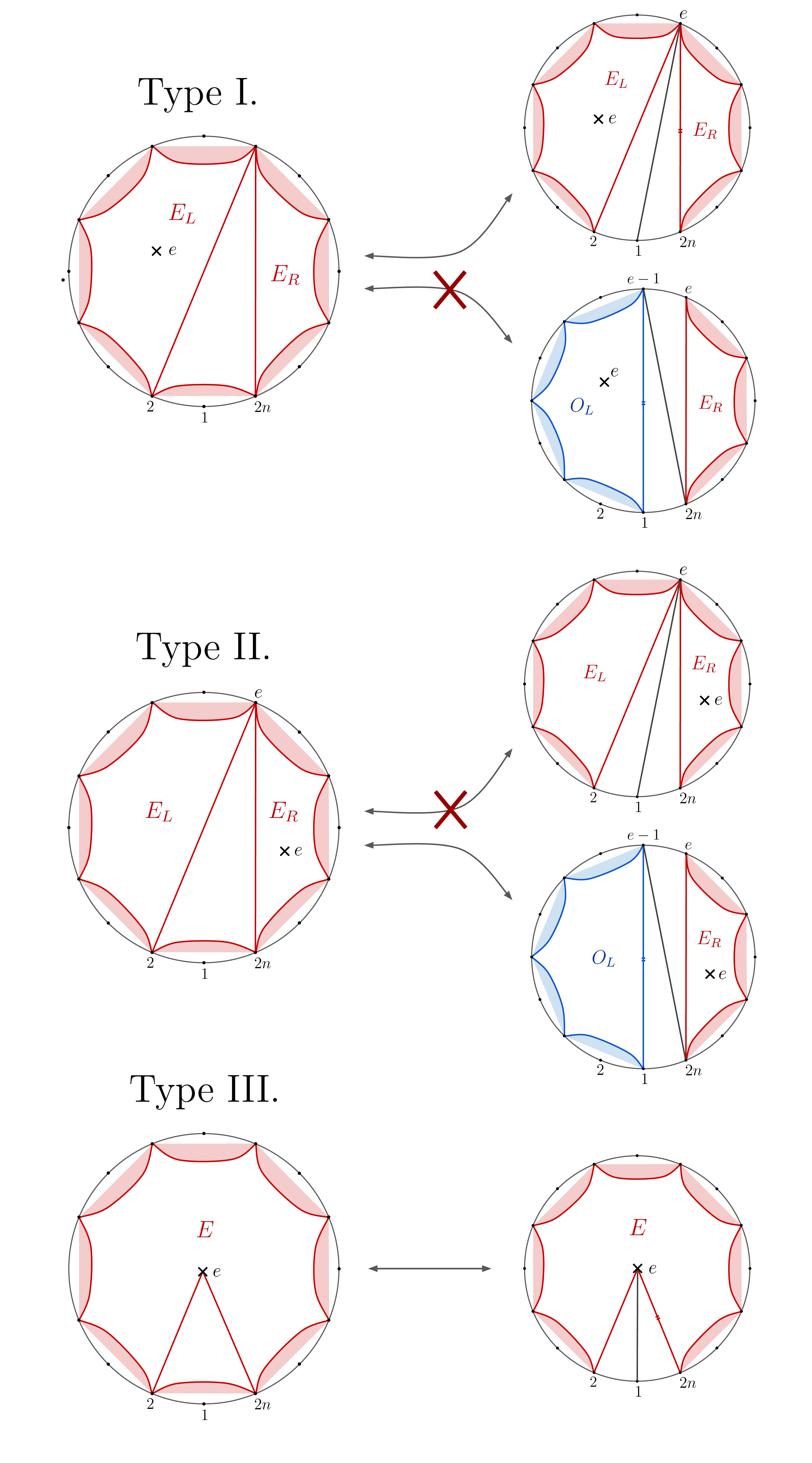}
    \caption{Bijections between contact terms and terms with single poles that survive the soft limit. With the puncture assigned to be even, we consider triangulations made with only $X_{e,e}$ chords on the left for the contact terms, indexed by the triangle $(2,2n,e)$ in which the chord $(2n,2)$ appears; in case III. $e$ is the puncture itself.  On the right, we indicate with crosses the triangulations which are irrelevant in the soft limit, because the presence of the puncture inside the even-gon picks up $+1$ instead of the requisite factor of $(-X)$. }
    \label{fig:LoopAdler}
\end{figure}

Now we will consider the three different possible types of triangles and explain how from $\alpha$ and $\beta$ we get contributions that precisely cancel the contact piece. This argument is summarized in a picture at the end of the note, fig. \ref{fig:LoopAdler}. Let's look at triangle type I. In this case, $\alpha_{2j,1}$ comes from the triangulations with a single $X_{o,e}$, $X_{2j,1}$, where we pick on the right problem, which is simply a tree-level problem, we pick the $X_{e,e} = X_{2j,2n}$, which automatically lands us on a pure $X_{e,e}$ triangulation for the right problem. On the left part, since it includes the puncture which is even, we can \textbf{only} have $X_{e,e}$ triangulations. Therefore $\alpha_{2j,1} = -N_{e,e}^L\times N_{e,e}^R$. Note that, as opposed to the tree-level, we don't get any contribution from $\beta_{2j-1,1}$, this is because from a triangulation containing $X_{2j-1,2n}$, we would need to pick the chord $X_{2j-1,1}$ from the left problem, to produce the correct numerator factor, and this would be selecting a pure $X_{o,o}$ triangulation on the left, which is impossible since it contains the puncture which is even. 

For a triangle of type II., the cancellation now comes from $\beta_{2n,2j-1}$. In this case, we start with a triangulation containing chord $X_{2n,2j-1}$, then on the left problem, which is now a tree problem, we need to pick chord $X_{1,2j-1}$ which picks an $X_{o,o}$ triangulation for the tree problem. On the right problem, since it contains the puncture we only get contributions from all the possible $X_{e,e}$ triangulations. Therefore we have that  $\beta_{2n,2j-1} = -N_{o,o}^L\times N_{e,e}^R =  -N_{e,e}^L\times N_{e,e}^R$. Now, once more, we don't have any contribution from $\alpha_{1,2j}$ because we need an $X_{1,2j}$ in the denominator together with an $X_{2n,2j}$ in the numerator, but $X_{2n,2j}$ is now part of the $X_{e,e}$ triangulation of the loop-right problem, and thus we only get a factor of $+1$, and not a term proportional to $X_{2n,2j}$. 

Let's now look at the triangle of type III., in this case, we only have contributions from $\alpha_{0,1}$ since the puncture parity is fixed to be even. For these terms, we have $X_{0,1}=Y_1$ in the denominator which must come with a $X_{0,2n}=Y_2$ in the numerator. Therefore we have $\alpha_{0,1s} = -N_{e,e}^{\text{tree}}$. 

The nature of this argument was essentially the same as what we saw at tree-level, and we expect similar logic to hold for establishing the Adler zero in the surface-soft limit for any number of loops. We have also checked explicitly that the integrand for the two-loop bubble vanishes in the surface-soft limit. 

\section{Tropical NLSM Amplitudes}

We now see how the shifted Tr$(\Phi^3)$ representation of the NLSM gives a natural and efficient tropical representation of NLSM amplitudes. In \cite{tropL}, a general procedure was given to build tropical representations of amplitudes for any Lagrangian for colored particles, $\mathcal{L}$, associated with the surface defining any order in the topological expansion, expressed in the form of a curve-integral \cite{curveint,curveint2}: 
\begin{equation}
{\cal A} = \int d^E t  \, {\rm exp} \, \left(-\sum_X \alpha_X X \right) \times {\cal N}_{\cal L},
\end{equation}
where the $\alpha_X(t)$ are the piecewise-linear ``headlight functions'' associated with curves $X$ on a surface, and $E$ is the number of propagators or equivalently the dimension of the Teichm{\"u}ller space of the surface. The curve integral is a ``global Schwinger parametrization''; this allows the Gaussian loop integrations to be performed leaving us with an integration over ${\bf t}$ space. In addition for general surfaces starting with the annulus at one-loop, we have to further mod out by the action of the mapping class group of the surface. This is accomplished by including the ``Mirzakhani Kernel'' ${\cal K}(\alpha)$ as a factor in the integrand. All these points are described in detail in \cite{curveint,curveint2}, but we suppress them here to focus on the novelties associated with the tropical representation of the NLSM.

The factor ${\cal N}_{\cal L}$ is a ``numerator function'' encoding the non-trivial interactions of ${\cal L}$. The action in the exponential has homogeneous weight in $t$ and so a single integration over an overall positive scaling can be performed to arrive at the ``Feynman'' form we present. One (arbitrary) choice is to restrict the integration to the unit sphere $S^{E-1}$, which gives the form 
\begin{equation}
{\cal A}_{\cal L} =  \int_{S^{E-1}}  \frac{\langle t d^{E-1} t \rangle \, {\cal N}_{{\cal L}}}{(\sum_X \alpha_X X)^{E}}.
\end{equation}

Other choices for fixing the overall rescaling on $t$ space are also natural; the choice of the sphere $S^{E-1}$ simply highlights that the integration can be taken over a compact space. 

The general procedure in \cite{tropL} for computing ${\cal N}_{\cal L}$ gives an individual prescription to construct each $m$-point interaction vertices of the Lagrangian. This provides an efficient, polynomial-time method to obtain amplitudes for Lagrangians with a finite number of terms, but becomes increasingly complicated for a theory like the NLSM with infinitely many vertices. 

The connection to Tr$(\Phi^3)$ offers a major simplification: we can simply take the Tr$(\Phi^3)$ curve integral and perform the kinematic shift $X_{e,e} \to X_{e,e} + \delta, X_{o,o} \to X_{o,o} - \delta$. Note that for the curve integrals to remain convergent after the shift, we should take $\delta$ to be imaginary. So we look at 
\begin{equation}
{\cal A}_{\delta} = \int_{S^{E-1}}  \frac{\langle t d^{E-1} t \rangle}{(\sum_X \alpha_X  X + \delta(\sum \alpha_{e,e} - \alpha_{o,o}))^{E}}.
\label{eq:dirshift}
\end{equation}

The $\alpha_X(t)$'s depend on a choice of reference triangulation defining the surface, and it is a nice fact that the object $\sum (\alpha_{e,e} - \alpha_{o,o})$ often simplifies greatly and becomes exactly (rather than piecewise) linear. For instance for any triangulations with no $(e,o)$ chords, we have that $\sum (\alpha_{e,e} - \alpha_{o,o}) = \sum_i t_i$. At loop level, we simply sum these expressions over all the parity assignments for the puncture. 

Now beginning with this with this expression, we can go one step further, and extract the leading term in the $\delta$ expansion to produce the two-derivative NLSM amplitude. Note that we can not naively Taylor expand the denominator in powers of $(X/\delta)$. The reason is obvious: the amplitude actually has poles in the $X_{e,o}$. This happens in the integral because there are regions in $t$ space where the coefficient $(\sum \alpha_{e,e} - \sum \alpha_{o,o})$ vanishes and so we can't expand in $1/\delta$. We can instead simply Taylor expand the integrand in $X_{e,e}, X_{o,o}$ since these do not occur as poles. 

But we know that the vast majority of terms in this Taylor expansion cancel out. For definiteness let's focus on the case of tree-amplitudes, where we know that only terms that are {\bf linear} in any $X_{e,e}, X_{o,o}$, with total weight $(N_{e,o} +1)$, survive in the final result. Here $N_{e,o}$ is the number of $(e,o)$ propagators, which we can easily detect via a tropical function using the $\Theta_X = {\bf g}_X \cdot \nabla \alpha_X$ variables introduced in~\cite{tropL}, which yield $\Theta_X(\Vec{t} \,\,) = 1$ for cones containing $X$ and $0$ otherwise. Thus
\begin{equation}
N_{e,o} = \sum_{(e,o)} \Theta_{X_{e,o}}.
\end{equation}

We can then extract the terms that are at most linear in $X_{e,e},X_{o,o}$ and of the correct weight by introducing a dummy variable $s$ and writing  
\begin{equation}
{\cal A}^{{\rm tree}}_{{\rm NLSM}} = \int_{S^{E-1}}  \langle t d^{E-1} t \rangle \times \frac{\left[\prod_{e,e} (1 - s \,\alpha_{X_{e,e}} X_{e,e}) \prod_{o,o} (1 + s \, \alpha_{X_{o,o}} X_{o,o} ) \right]_{N_{e,o} + 1}}{(\sum_{e,o} \alpha_{X_{e,o}}  X_{e,o} + \delta(\sum \alpha_{e,e} - \alpha_{o,o}))^{E + N_{e,o} + 1 }} .
\end{equation}

The coefficient of $s^{N_{e,o} + 1}$ from the product gives monomials of weight $N_{e,o} + 1$ that are linear in each $X_{e,e},X_{o,o}$, exactly matches the coefficients of the same monomials from the direct Taylor expansion of equation (\ref{eq:dirshift}) in $X_{e,e},X_{o,o}$, exactly as needed.

This tropical expression differs significantly from the tropical formulae for general Lagrangians of \cite{tropL}, not least because even the denominator structure is different, with the $\delta$ deformation. Another crucial difference is that now the numerator factors depend explicitly on the $\alpha$'s and not just on $\Theta$'s. We can give a different tropical formulation starting directly from the numerator factors discovered for shifted Tr$(\Phi^3)$, where all the terms that cancel in pairs in the sum over triangulations have been eliminated. Here we have simply 
\begin{equation}
{\cal A}^{{\rm tree}}_{{\rm NLSM}} = \int_{S^{E-1}}  \langle t d^{E-1} t \rangle \times\frac{ (-1)^{N_{o,o}}  \left[\prod_{e,e} (1 - s \Theta_{X_{e,e}} X_{e,e}) \prod_{o,o} (1 +  s \Theta_{X_{o,o}} X_{o,o}) \right]_{N_{e,o} + 1}} {(\sum_{e,o} \alpha_{X_{e,o}}  X_{e,o} + \sum_{e,e} \alpha_{e,e} + \sum_{o,o} \alpha_{o,o})^{E}},  
\label{eq:trop2}
\end{equation}
with $N_{o,o}$ defined analogously to $N_{e,o}$ as $N_{o,o} = \sum_{o,o} \Theta_{X_{o,o}}$. Note that in the denominator we have set the coefficients of $\alpha_{e,e}, \alpha_{o,o} \to 1$, so that the denominators from each cone just give us the $1/X_{e,o}$ poles. The numerator is simply extracting all monomials in $(-X_{e,e})$ and $(+ X_{o,o})$ of total weight $N_{e,o} + 1$, multiplied by the overall sign $(-1)^{N_{e,o} + 1}$. 

We have found very simple tropical representations for NLSM amplitudes using the numerator functions most naturally associated with the shifted Tr$(\Phi^3)$ amplitudes, rather than the one more directly connected to the NLSM via the Catalan rule. The reason is simple, the Catalan rule needs to ``detect'' rather intricate and detailed features of the triangulation: the even-gons produced by the $(e,o)$ chords, and the $(e,e),(o,o)$ chords triangulating their interiors. By contrast, the shifted Tr$(\Phi^3)$ rules need much ``coarser'' information about the triangulation, with the numerators simply consisting of the product of $X_{e,e}, X_{o,o}$'s present in the triangulation with a fixed total weight. 

Let's now turn to the tropical representation at loop level. The first point to make is that obviously, loop integration in the pure NLSM will give us divergences, that will have to be regulated if we first extract the leading NLSM amplitudes from the shifted Tr$(\Phi^3)$ theory. In this regard of course it is very interesting that the Tr$(\Phi^3)$ theory is perfectly healthy in the UV in $D \leq 4$ dimensions, with only power divergences associated with the tadpoles that are trivial to excise tropically as described in \cite{curveint}. So long as we perform the $\delta$ deformation with pure imaginary $\delta$, all the loop integrals will be finite, as the shifted Tr($\Phi^3$) theory provides an unusual UV regularization of the NLSM. 

It is nonetheless still of interest to represent the ``pure'' NLSM integrand tropically, and this is what we now do. We begin at one-loop. Again we assign a parity to the puncture and sum our expressions over both parities. As we saw the rule for the cones for which the $(e,o)$ chords touch the puncture is exactly the same as at tree-level; the only difference is that the total weight in the $X_{e,e}, X_{o,o}$ is just $N_{e,o}$ rather than $(N_{e,o} + 1)$ as at tree-level. But we also have to make sure to weight the cones for which the $(e,o)$ chords don't touch the puncture with a factor of $0$, (or assign a factor of $+1$ to the corresponding even-gon for the shifted Tr($\Phi^3$) integrand, but it will be easier to enforce the vanishing weight). But this is trivial to detect. In terms where the puncture is taken to be even/odd $p_{e/o}$, we simply include an additional factor 
\begin{equation}
{\cal P}_{e/o} = 1 - \prod_{o/e} (1 - \Theta_{X_{o/e,p_{e/o}}}),
\end{equation}
which is 1 in any cone that contains at least a single $(o,p_e)$ chord, and zero otherwise. In this way the analogue of equation (\ref{eq:trop2}) for the 1-loop NLSM integrand is 
\begin{equation}
{\cal A}^{{\rm one-loop}}_{{\rm NLSM}} = \sum_{p_{e/o}} \int_{S^{E-1}}  \langle t d^{E-1} t \rangle   \frac{(-1)^{N_{o,o}}{\cal P}_{e/o} \left[\prod_{e,e} (1 - s \Theta_{X_{e,e}} X_{e,e}) \prod_{o,o} (1 +  s \Theta_{X_{o,o}} X_{o,o}) \right]_{N_{e,o}}} {(\sum_{e,o} \alpha_{X_{e,o}}  X_{e,o} + \sum_{e,e} \alpha_{e,e} + \sum_{o,o} \alpha_{o,o})^{E}}.
\end{equation}

The extension to general $L$-loop order in the planar limit is straightforward. For those cones where all punctures are touched by $(e,o)$ chords, we simply pick out the contribution that has weight $(N_{e,o} + L - 1)$, and we have to include a factor ${\cal P}_{e/o}$ for every puncture to give all the other cones vanishing weight. 

\section{Outlook}

The existence of a simple and direct relationship between Tr$(\Phi^3)$ and the NLSM is at first quite surprising -- why should the theory with the simplest, non-derivative interactions have anything whatsoever to do with a very special derivatively coupled theory with an infinite tower of even-point interactions? 

The associahedron/stringy representations of Tr$(\Phi^3)$ amplitudes provide what still appears to be the simplest and deepest conceptual understanding of this connection, making many further features of the amplitudes obvious at the same time. Nonetheless in this note, we have highlighted a different and more elementary but still conceptually interesting reason for the connection beginning with the NLSM itself. This opens up a number of obvious questions that would be worth understanding better, three of which we mention here in closing. 
\\ \\
\noindent{\bf Multisoft limits}: The stringy formulation of Tr$(\Phi^3)$ and the kinematic shift to the NLSM makes the Adler zero for single-soft emission obvious. Beyond single-soft pions, the NLSM has long been expected to have an interesting factorization pattern for multi-soft limits. This has been well-understood for double-soft limits \cite{simplest}, as the momenta $p^\mu_{2n-1}, p^\mu_{2n} \to 0$. We can again phrase this limit cleanly using the $X_{i,j}$ kinematic variables, as an obvious generalization of the ``surface-soft'' limit to multiple pions. Now the limit is associated with putting the external points $\{1, 2n-1, 2n\}$ on top of each other to get a smaller surface. In doing this, some of the curves $C$ of the initial surface become identified with curves $\tilde{C}$ of the smaller surface. If $\tilde{C}$ is an interior curve of the smaller surface, we simply map $X_C \to X_{\tilde{C}}$. But if $\tilde{C}$ is a boundary curve, we have the on-shell condition that $X_{\tilde{C}} \to 0$, and so we should send $X_{C} \to 0$. However in the multisoft limit we anticipate $0/0$ factors from taking momenta soft, and so we will send these $X_C \to 0$ carefully, by first scaling $X_C \to t X_C$ and then sending $t \to 0$. 

For instance working at tree-level, we simply send $\{X_{j,2n}, X_{j,2n-1}\} \to X_{1,j}$ for $j=3,\cdots, 2n-3$, and scale 
$X_{a,b} \to t X_{a,b}$ with $t \to 0$, for $a,b \in \{1,2,2n-2,2n-1\}$. The amplitude doesn't vanish in this double-soft limit due to $t/t$ factors as $t \to 0$. However, it factorizes beautifully as 
\begin{equation}
\begin{aligned}
 {\cal A}^{{\rm 2\, soft}}_{{\rm NLSM}, 2n} \to  {\cal A}_{{\rm NLSM}, 2n - 2} \times  \left(1 - \frac{X_{1,2n-1} + X_{2,2n}}{X_{2,2n-1}} - \frac{X_{1,2n-1} + X_{2n-2,2n}}{X_{1,2n-2}} \right). 
\end{aligned}
\end{equation}

This way of defining the soft limit by $t$ rescalings of $X$ variables is most natural from the surface perspective and since the $X$'s define a kinematic basis, also gives a perfectly well-defined way of taking a soft limit in terms of momenta. But it is interesting to compare this definition with the more conventional expectation coming from rescaling momenta as $p^\mu_{2n} \to s \,p_{2n}^\mu, p^\mu_{2n-1} \to s\, p^\mu_{2n -1}$, sending $s \to 0$. In this limit, we have that $X_{a,b} \to s X_{a,b}$ for $a,b \in \{1,2,2n-2,2n-1\}$ {\it except} for $X_{1,2n-1} = (p_{2n} + p_{2n-1})^2 \to s^2 X_{1,2n-1}$. Our $t$ scaling is thus seen to be more general, containing the momentum $s$ scaling as a special case, since in the $s$-scaling we clearly have that $X_{1,2n-1} \to 0$ relative to the other $X$'s and so we simply set $X_{1,2n-1} \to 0$ in the soft factor. But the full object with the $t$ scaling is both more general and more natural. Indeed, the double-soft factor given above -- with $X_{1,2n-1} \neq 0$ -- can be interpreted as a mixed amplitude for two pions and three $\phi$'s; this understanding would be obscured by setting $X_{1,2n-1} \to 0$. In a moment we will see how this generalizes to multi-soft limits. 

The double-soft limit of pion amplitudes has been well-studied and derived from a number of points of view \cite{simplest,SingleDouble,DoubleSoftShift,JaraNLSM}. As explained in \cite{simplest}, the double-soft behavior is exploring the $[U(N) \times U(N)]/U(N)$ coset around the identity at second-order, reflecting the commutation relation $[X,X] \sim T$ of two broken generators corresponding to the soft pions giving an unbroken one. The multi-soft pion limits should fully explore the coset space. The soft limit for an odd number of pions vanishes, but we expect a simple result for an even number $(2m)$ of adjacent pions, with \cite{multisoft} giving a form of the result using recursive techniques at tree-level. 

As will be shown in \cite{upcoming}, there is an essentially one-line conceptual understanding of these limits, following from a much more general pattern of factorizations for stringy Tr$(\Phi^3)$ amplitudes at all loop orders, making use of the simple but deep fact that the binary $u$ variables for a subsurface $s$ of a surface $S$ can be expressed as monomials of the $u$ variables of $S$. After the $\delta$ shifts are performed to get NLSM amplitudes, these general expressions reduce to a beautiful interpretation for the soft factor in the limit where the momenta $2n - 2m, \cdots, 2n$ are taken soft. In terms of $X_{i,j}$ this is the scaling where 
\begin{equation} 
\begin{cases}
    X_{j,a} \to X_{j,1} \text{ and }  X_{a,j} \to X_{1,j} , \quad {\rm for} \, j \in \{3,4,\cdots, 2n - m - 1\}, \\
    X_{a,b} \to t X_{a,b}, \,  t \to 0, \quad  {\rm for} \, a,b \in \{1,2,2n-2m, 2n - 2m + 1, \cdots, 2n\} .
\end{cases} 
\end{equation}

\begin{figure}[t]
    \centering
    \includegraphics[width=\textwidth]{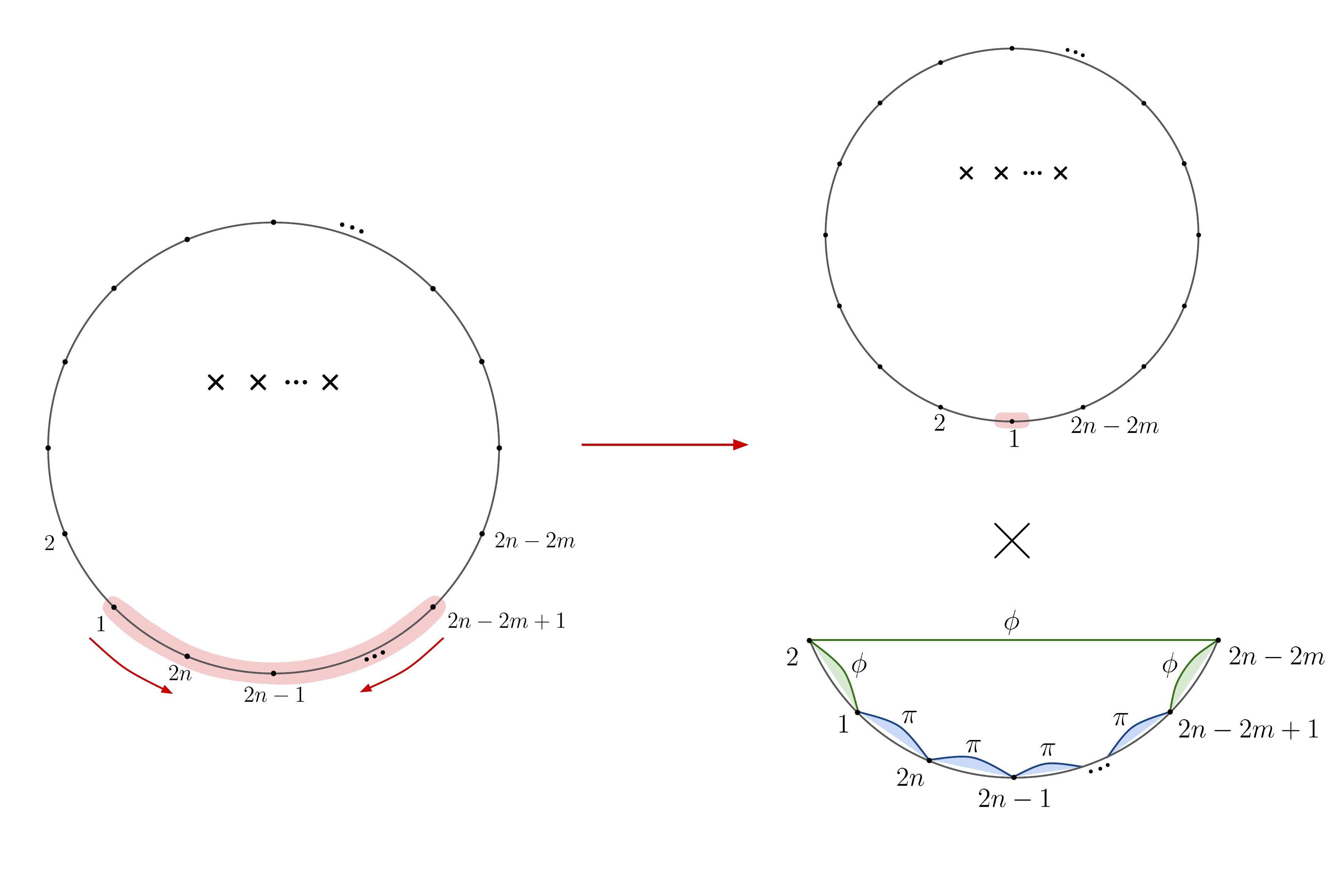}
    \caption{Tree amplitudes and loop integrands for the shifted Tr ($\Phi^3)$ theory, and hence also for the NLSM, have a simple universal behavior as any even number $2m$ of adjacent pions are made ``surface soft'', factorizing into the $2(n-m)$ point integrand multiplied by a soft factor given by a tree-level ``mixed'' amplitude with $2m$ $\pi$'s and 3 $\phi$'s.}
    \label{fig:multisoft}
\end{figure}
In this limit, the amplitude factorizes into the lower $(2n - 2m)$ pion amplitude, multiplied by the mixed amplitude for $2m$ pions plus $3$ $\Phi$'s,  as in fig. \ref{fig:multisoft}:
\begin{equation}
{\cal A}^{2m-{\rm soft}}_{{\rm NLSM}, 2n} \to {\cal A}_{{\rm NLSM}, 2n - 2m} \times {\cal A}_{\Phi \Phi \Phi \pi \cdots \pi}(1,2,2n-2m,\cdots, 2n) ,
\end{equation}
where in turn this mixed amplitude is trivially obtained from the shifted Tr$(\Phi^3)$ amplitude as 
\begin{equation}
 {\cal A}_{\Phi \Phi \Phi \pi \cdots \pi} ={\rm lim_{\delta \to \infty}} \delta^{2 m} {\cal A}^{{\rm tree}}_{{\rm Tr}(\Phi^3)}(1,2,2n-2m, \cdots, 2n)\big\vert_{X_{e,e/o,o} \to X_{e,e/o,o} \pm \delta}.
\end{equation}

More generally the NLSM amplitude above can be replaced by the fully shifted Tr$(\Phi^3)$ amplitude working at finite $\delta$. The soft limit when even groups of pions are separated simply factor into the product of the $2m$ adjacent ones we have just given, while if any odd group of pions is isolated we get zero. 

The same result holds for loop integrands, using the surface-multsoft limits. The $2m$-soft factors are still given by the tree-level result, while the NLSM factor can be replaced by the shifted Tr$(\Phi^3)$ integrand at any finite $\delta$. Remarkably, as we also saw for the Adler zero, these factorizations do {\it not} require summing over the different parity assignments for the punctures, they hold for each such assignment individually. 

In \cite{upcoming} we will see how these results are made entirely obvious from the binary geometry magic of ``surfaceology''. Nonetheless, the final results are so simple and striking that they beg for a direct combinatorial/graphical understanding along the lines of the understanding for the Adler zero at tree- and loop-level we have seen in this note.
\\ \\
\noindent {\bf Large chiral logs}: The study of the NLSM was historically extremely important for developing our modern understanding of effective field theory, in a setting beyond the rather more trivial dimensional analysis reflected in the division between relevant, marginal and irrelevant couplings.  As emphasized by Weinberg in his earliest papers on the subject, while the NLSM is a non-renormalizable theory, the logarithmic divergences in the theory induce entirely calculable higher-dimension operators. In position space, these are associated with small but power-law corrections, in principle detectable by long-distance measurements,  as opposed to the incalculable contact terms that are only important at short distances. In Euclidean momentum space, the logs are calculable because they receive contributions equally from all scales; in Lorentzian signature the logs are calculable because their cuts are computed by on-shell amplitudes for the low-energy pions. Now in the curve-integral formalism the logarithmic divergences can be detected in the tropical fan, as sketched in \cite{curveint}; it would be interesting to adapt this to the tropical representations we have given for the NLSM, either by leaving $\delta$ as a UV regulator in the shifted Tr$(\Phi^3)$ formulae or after extracting the pure NLSM from the leading terms in the $\delta$ expansion, to give a tropical representation for the leading log amplitudes in the theory at any multiplicity. 
\\ \\
\noindent {\bf Good triangulation representations for Yang-Mills}: Another interesting lesson of our analysis is the importance of finding the ``right'' representation of amplitudes for a given theory as a sum over numerator-dressed cubic graphs/triangulations. Of course, such representations are not unique -- higher point vertices can be blown up into cubic graphs in many different ways -- but we found an especially natural one for the NLSM. Now we have already commented on the remarkable extension of shifted kinematics to describe gluon amplitudes beginning with the shifted Tr$(\Phi^3)$ theory~\cite{gluons}. But could there be a different, similarly ``right'' sum over triangulations picture for Yang-Mills amplitudes already in the field theory limit, that would allow the correct understanding of its amplitudes as well as tropical representations for them, in a way analogous to what we have seen in this note for the NLSM?

\acknowledgments 
We thank Christoph Bartsch, Song He, Karol Kampf, Jiri Novotny, and Jaroslav Trnka for useful discussions.
N.A.H. is supported by the DOE (Grant No.~DE-SC0009988), by the Simons Collaboration on Celestial Holography, and further support was made possible
by the Carl B. Feinberg cross-disciplinary program in innovation at the IAS. C.F. is supported by FCT/Portugal (Grant No.~2023.01221.BD).
\bibliographystyle{JHEP}\bibliography{Refs}

\providecommand{\href}[2]{#2}\begingroup\raggedright\begin{thebibliography}{10}

\bibitem{RecFromSoftTheorems}
H.~Luo and C.~Wen, {\it {Recursion relations from soft theorems}},  {\em JHEP}
  {\bf 03} (2016) 088, [\href{http://arxiv.org/abs/1512.06801}{{\tt
  arXiv:1512.06801}}].

\bibitem{RecursionJara}
K.~Kampf, J.~Novotny, and J.~Trnka, {\it {Recursion relations for tree-level
  amplitudes in the $SU(N)$ nonlinear sigma model}},  {\em Phys. Rev. D} {\bf
  87} (2013), no.~8 081701, [\href{http://arxiv.org/abs/1212.5224}{{\tt
  arXiv:1212.5224}}].

\bibitem{OnShellRecRel}
C.~Cheung, K.~Kampf, J.~Novotny, C.-H. Shen, and J.~Trnka, {\it {On-Shell
  Recursion Relations for Effective Field Theories}},  {\em Phys. Rev. Lett.}
  {\bf 116} (2016), no.~4 041601, [\href{http://arxiv.org/abs/1509.03309}{{\tt
  arXiv:1509.03309}}].

\bibitem{AllEFTsCliffJara}
C.~Cheung, K.~Kampf, J.~Novotny, C.-H. Shen, and J.~Trnka, {\it {A Periodic
  Table of Effective Field Theories}},  {\em JHEP} {\bf 02} (2017) 020,
  [\href{http://arxiv.org/abs/1611.03137}{{\tt arXiv:1611.03137}}].

\bibitem{DoubleCopyAndSoft}
K.~Zhou and F.-S. Wei, {\it {Note on NLSM tree amplitudes and soft theorems}},
  {\em Eur. Phys. J. C} {\bf 84} (2024), no.~1 68,
  [\href{http://arxiv.org/abs/2306.09733}{{\tt arXiv:2306.09733}}].

\bibitem{EFTsSoftLimits}
C.~Cheung, K.~Kampf, J.~Novotny, and J.~Trnka, {\it {Effective Field Theories
  from Soft Limits of Scattering Amplitudes}},  {\em Phys. Rev. Lett.} {\bf
  114} (2015), no.~22 221602, [\href{http://arxiv.org/abs/1412.4095}{{\tt
  arXiv:1412.4095}}].

\bibitem{ScalarBCJBoostrap}
T.~V. Brown, K.~Kampf, U.~Oktem, S.~Paranjape, and J.~Trnka, {\it {Scalar
  Bern-Carrasco-Johansson bootstrap}},  {\em Phys. Rev. D} {\bf 108} (2023),
  no.~10 105008, [\href{http://arxiv.org/abs/2305.05688}{{\tt
  arXiv:2305.05688}}].

\bibitem{softBootstrap}
I.~Low and Z.~Yin, {\it {Soft Bootstrap and Effective Field Theories}},  {\em
  JHEP} {\bf 11} (2019) 078, [\href{http://arxiv.org/abs/1904.12859}{{\tt
  arXiv:1904.12859}}].

\bibitem{Uniqueness2}
L.~Rodina, {\it {Uniqueness from gauge invariance and the Adler zero}},  {\em
  JHEP} {\bf 09} (2019) 084, [\href{http://arxiv.org/abs/1612.06342}{{\tt
  arXiv:1612.06342}}].

\bibitem{UniquenessNLSM}
N.~Arkani-Hamed, L.~Rodina, and J.~Trnka, {\it {Locality and Unitarity of
  Scattering Amplitudes from Singularities and Gauge Invariance}},  {\em Phys.
  Rev. Lett.} {\bf 120} (2018), no.~23 231602,
  [\href{http://arxiv.org/abs/1612.02797}{{\tt arXiv:1612.02797}}].

\bibitem{UVconisderations}
J.~J.~M. Carrasco and L.~Rodina, {\it {UV considerations on scattering
  amplitudes in a web of theories}},  {\em Phys. Rev. D} {\bf 100} (2019),
  no.~12 125007, [\href{http://arxiv.org/abs/1908.08033}{{\tt
  arXiv:1908.08033}}].

\bibitem{CHY}
F.~Cachazo, S.~He, and E.~Y. Yuan, {\it {Scattering Equations and Matrices:
  From Einstein To Yang-Mills, DBI and NLSM}},  {\em JHEP} {\bf 07} (2015) 149,
  [\href{http://arxiv.org/abs/1412.3479}{{\tt arXiv:1412.3479}}].

\bibitem{FlavorKinRepresentation}
C.~Cheung and C.-H. Shen, {\it {Symmetry for Flavor-Kinematics Duality from an
  Action}},  {\em Phys. Rev. Lett.} {\bf 118} (2017), no.~12 121601,
  [\href{http://arxiv.org/abs/1612.00868}{{\tt arXiv:1612.00868}}].

\bibitem{zeros}
N.~Arkani-Hamed, Q.~Cao, J.~Dong, C.~Figueiredo, and S.~He, {\it {Hidden zeros
  for particle/string amplitudes and the unity of colored scalars, pions and
  gluons}},  \href{http://arxiv.org/abs/2312.16282}{{\tt arXiv:2312.16282}}.

\bibitem{deltaNLSM}
N.~Arkani-Hamed, Q.~Cao, J.~Dong, C.~Figueiredo, and S.~He, {\it {NLSM
  $\subset$ Tr$(\phi^3)$}},  \href{http://arxiv.org/abs/2401.05483}{{\tt
  arXiv:2401.05483}}.

\bibitem{ABHY}
N.~Arkani-Hamed, Y.~Bai, S.~He, and G.~Yan, {\it {Scattering Forms and the
  Positive Geometry of Kinematics, Color and the Worldsheet}},  {\em JHEP} {\bf
  05} (2018) 096, [\href{http://arxiv.org/abs/1711.09102}{{\tt
  arXiv:1711.09102}}].

\bibitem{GiulioClusters}
N.~Arkani-Hamed, S.~He, G.~Salvatori, and H.~Thomas, {\it {Causal diamonds,
  cluster polytopes and scattering amplitudes}},  {\em JHEP} {\bf 11} (2022)
  049, [\href{http://arxiv.org/abs/1912.12948}{{\tt arXiv:1912.12948}}].

\bibitem{StringForms}
N.~Arkani-Hamed, S.~He, T.~Lam, and H.~Thomas, {\it {Binary geometries,
  generalized particles and strings, and cluster algebras}},  {\em Phys. Rev.
  D} {\bf 107} (2023), no.~6 066015,
  [\href{http://arxiv.org/abs/1912.11764}{{\tt arXiv:1912.11764}}].

\bibitem{StringyCanonicalForms}
N.~Arkani-Hamed, S.~He, and T.~Lam, {\it {Stringy canonical forms}},  {\em
  JHEP} {\bf 02} (2021) 069, [\href{http://arxiv.org/abs/1912.08707}{{\tt
  arXiv:1912.08707}}].

\bibitem{WignerSemiCircle}
E.~P. Wigner, {\it On the distribution of the roots of certain symmetric
  matrices},  {\em Annals of Mathematics} {\bf 67} (1958), no.~2 325--327.

\bibitem{AdlerZero}
S.~L. Adler, {\it Consistency conditions on the strong interactions implied by
  a partially conserved axial-vector current},  {\em Phys. Rev.} {\bf 137}
  (Feb, 1965) B1022--B1033.

\bibitem{JaraSoftTheorem}
C.~Bartsch, K.~Kampf, J.~Novotny, and J.~Trnka, {\it {An All-loop Soft Theorem
  for Pions}},  \href{http://arxiv.org/abs/2401.04731}{{\tt arXiv:2401.04731}}.

\bibitem{upcoming}
{\it {work in progress}},  \href{http://arxiv.org/abs/24xx.xxxx}{{\tt
  arXiv:24xx.xxxx}}.

\bibitem{JaraNLSM}
K.~Kampf, J.~Novotny, and J.~Trnka, {\it {Tree-level Amplitudes in the
  Nonlinear Sigma Model}},  {\em JHEP} {\bf 05} (2013) 032,
  [\href{http://arxiv.org/abs/1304.3048}{{\tt arXiv:1304.3048}}].

\bibitem{cubicnum1}
S.~He, L.~Hou, J.~Tian, and Y.~Zhang, {\it {Kinematic numerators from the
  worldsheet: cubic trees from labelled trees}},  {\em JHEP} {\bf 08} (2021)
  118, [\href{http://arxiv.org/abs/2103.15810}{{\tt arXiv:2103.15810}}].
  [Erratum: JHEP 06, 037 (2022)].

\bibitem{cubicnum2}
C.~Cheung and J.~Mangan, {\it {Covariant color-kinematics duality}},  {\em
  JHEP} {\bf 11} (2021) 069, [\href{http://arxiv.org/abs/2108.02276}{{\tt
  arXiv:2108.02276}}].

\bibitem{cubicnum3}
J.~J.~M. Carrasco, C.~R. Mafra, and O.~Schlotterer, {\it {Abelian Z-theory:
  NLSM amplitudes and $\alpha$'-corrections from the open string}},  {\em JHEP}
  {\bf 06} (2017) 093, [\href{http://arxiv.org/abs/1608.02569}{{\tt
  arXiv:1608.02569}}].

\bibitem{tropL}
N.~Arkani-Hamed, C.~Figueiredo, H.~Frost, and G.~Salvatori, {\it {Tropical
  Amplitudes For Colored Lagrangians}},
  \href{http://arxiv.org/abs/2402.06719}{{\tt arXiv:2402.06719}}.

\bibitem{LoopRecursionJara}
C.~Bartsch, K.~Kampf, and J.~Trnka, {\it {Recursion relations for one-loop
  Goldstone boson amplitudes}},  {\em Phys. Rev. D} {\bf 106} (2022), no.~7
  076008, [\href{http://arxiv.org/abs/2206.04694}{{\tt arXiv:2206.04694}}].

\bibitem{curveint}
N.~Arkani-Hamed, H.~Frost, G.~Salvatori, P.-G. Plamondon, and H.~Thomas, {\it
  {All Loop Scattering as a Counting Problem}},
  \href{http://arxiv.org/abs/2309.15913}{{\tt arXiv:2309.15913}}.

\bibitem{curveint2}
N.~Arkani-Hamed, H.~Frost, G.~Salvatori, P.-G. Plamondon, and H.~Thomas, {\it
  {All Loop Scattering For All Multiplicity}},
  \href{http://arxiv.org/abs/2311.09284}{{\tt arXiv:2311.09284}}.

\bibitem{simplest}
N.~Arkani-Hamed, F.~Cachazo, and J.~Kaplan, {\it {What is the Simplest Quantum
  Field Theory?}},  {\em JHEP} {\bf 09} (2010) 016,
  [\href{http://arxiv.org/abs/0808.1446}{{\tt arXiv:0808.1446}}].

\bibitem{SingleDouble}
Y.-J. Du and H.~Luo, {\it {On single and double soft behaviors in NLSM}},  {\em
  JHEP} {\bf 08} (2015) 058, [\href{http://arxiv.org/abs/1505.04411}{{\tt
  arXiv:1505.04411}}].

\bibitem{DoubleSoftShift}
I.~Low, {\it {Double Soft Theorems and Shift Symmetry in Nonlinear Sigma
  Models}},  {\em Phys. Rev. D} {\bf 93} (2016), no.~4 045032,
  [\href{http://arxiv.org/abs/1512.01232}{{\tt arXiv:1512.01232}}].

\bibitem{multisoft}
Y.-J. Du and H.~Luo, {\it {Leading order multi-soft behaviors of tree
  amplitudes in NLSM}},  {\em JHEP} {\bf 03} (2017) 062,
  [\href{http://arxiv.org/abs/1611.07479}{{\tt arXiv:1611.07479}}].

\bibitem{gluons}
N.~Arkani-Hamed, Q.~Cao, J.~Dong, C.~Figueiredo, and S.~He, {\it
  {Scalar-Scaffolded Gluons and the Combinatorial Origins of Yang-Mills
  Theory}},  \href{http://arxiv.org/abs/2401.00041}{{\tt arXiv:2401.00041}}.

\end{thebibliography}\endgroup
\end{document}